\documentclass[%
 reprint,
superscriptaddress,
 amsmath,amssymb,
 nofootinbib,
 aps,
 prd,
 groupedaddress
]{revtex4-2}

\usepackage[margin=0.75in]{geometry}

\makeatletter
\def\ps@pprintTitle{%
\let\@oddhead\@empty
\let\@evenhead\@empty
\let\@oddfoot\@empty
\let\@evenfoot\@oddfoot
}
\makeatother

\usepackage{amssymb,amsmath}
\usepackage{mathrsfs}
\usepackage{graphicx}%
\usepackage{dcolumn}%
\usepackage{bm}%
\usepackage{subcaption}
\usepackage{float}
\usepackage{caption}
\usepackage{lipsum}  
\usepackage{tabularx}
\usepackage{ragged2e}
\usepackage{graphicx}
\usepackage{upgreek}  % in preamble
\usepackage{subcaption}
\usepackage{microtype}

\usepackage{caption}
\DeclareCaptionFont{justified}{\justifying}

\usepackage{xcolor}

\newcommand{\trento}{\texttt{ T$\mathrel{\protect\raisebox{-2.1pt}{R}}$ENT$\mathrel{\protect\raisebox{0pt}{o}}$}}

\newcommand{\la}{\langle}
\newcommand{\ra}{\rangle}

\begin{document}

\title{
Nonconformal kinetic theory with transverse dynamics and event-by-event simulations using \texttt{CoMBolt-RTA}
}%

\author{Seyed Farid Taghavi}
\affiliation{TUM School of Natural Sciences, Technische Universit\"at M\"unchen, Garching, Germany}%

\begin{abstract}

A new version of \texttt{CoMBolt}, called \texttt{CoMBolt-RTA}, is introduced to solve the Boltzmann equation for massive quasiparticles in the relaxation time approximation (RTA). The inclusion of a constant quasiparticle mass allows one to obtain an equation of state (EOS) closer to that of lattice QCD. The nonequilibrium dynamics of a boost-invariant, nonconformal system with transverse spatial extension are studied, with particular emphasis on the effects of quasiparticle mass on the evolution toward equilibrium. Using the hybrid version, which includes \trento{} for the initial state and \texttt{UrQMD} as an afterburner, the model parameters are manually tuned to describe recent anisotropic-flow and charged-multiplicity measurements in OO collisions by the ALICE Collaboration. The resulting predictions for the $p_T$-spectrum and $\langle p_T\rangle$ are also compared with ATLAS measurements. Based on the preferred parameter tuning, the collective response as a function of opacity is investigated. The results suggest that central OO collisions remain in the fluid-like regime, while smaller systems, namely ${}^3$He--${}^3$He and ${}^4$He--${}^4$He collisions, could provide a better opportunity to investigate the transition from fluid-like to particle-like collective dynamics.

\end{abstract}

\maketitle

%\tableofcontents
% \addtocontents{toc}{\protect\setcounter{tocdepth}{1}}
%     \addcontentsline{file}{section_unit}{entry}

\section{Introduction}

Collectivity is one of the main signatures of the production of hot QCD matter in the deconfined phase, the Quark-Gluon Plasma (QGP), in high-energy nuclear collisions. In the past decades, the observation of anisotropic flow in large-ion collisions such as Au--Au at RHIC~\cite{STAR:2000ekf} and Pb--Pb at the LHC~\cite{ALICE:2010suc} has been successfully modeled using relativistic causal hydrodynamics~\cite{Heinz:2024jwu}. The observation of collective signatures in small-system collisions, namely pp and p--Pb~\cite{CMS:2010ifv,ALICE:2012eyl,ATLAS:2012cix}, has raised a long-standing debate about the nature of the observed signals and the applicability of hydrodynamics. Recently, system-size scans and light-ion collisions have opened the opportunity to better understand the nature of the observed signals~\cite{ALICE:2025luc,ATLAS:2025nnt,CMS:2025tga}.
 
Using a kinetic or transport picture to describe the evolution of the QGP has a long history in the heavy-ion physics community. For instance, transport models such as AMPT~\cite{Lin:2004en} and BAMPS~\cite{Xu:2007aa} describe microscopic interactions among partonic degrees of freedom. Although one may question the applicability of this particle-based description, which is better justified for weakly coupled systems, these models have successfully described various collective-flow observables. In AMPT, the partonic cross-section is adjusted phenomenologically, while BAMPS employs scattering processes based on perturbative QCD. Relativistic transport models using geometric collision criteria can also exhibit collision-scheme dependence and causality violations, which need to be treated carefully~\cite{Zhang:1996gb}.

In recent years, using a quasiparticle picture and an approach closer to effective kinetic theory (EKT) has become more popular~\cite{Romatschke:2011hm,Kurkela:2017xis,Kurkela:2019kip,Kurkela:2020wwb,Ambrus:2021fej,Nugara:2024net}. In this approach, the Boltzmann equation is directly solved, and then the energy-momentum tensor is obtained from the underlying evolving Boltzmann distribution. Most of these models employ a simplified collision kernel, typically within RTA. Using this approach, one can start the evolution from far-from-equilibrium conditions and cover the full range of dynamics from free-streaming to ideal hydrodynamics, including regimes beyond the applicability of conventional methods based on hydrodynamics.

The current work follows the previous attempt to develop a realistic event generator called \texttt{CoMBolt}. The advancement compared to the previous version in Refs.~\cite{Taghavi:2025xhl,Taghavi:2025ddm,CoMBoltGithub} is that the Boltzmann equation is solved for massive quasiparticles. The new version is called \texttt{CoMBolt-RTA} to distinguish it from the previous massless version, \texttt{CoMBolt-ITA}, where ITA stands for isotropization time approximation. This allows one to have a more realistic EOS, which is important for a more realistic $p_T$-spectrum and $\la p_T\ra$. In this new version, a constant mass is considered, which means that an input EOS still cannot be employed. The temperature-dependent mass, as discussed in Ref.~\cite{Romatschke:2011qp}, still requires more effort to be implemented in the future.

\begin{figure*}[t!]
	\centering
	\begin{tabular}{cc}
		\includegraphics[width=0.5\textwidth]{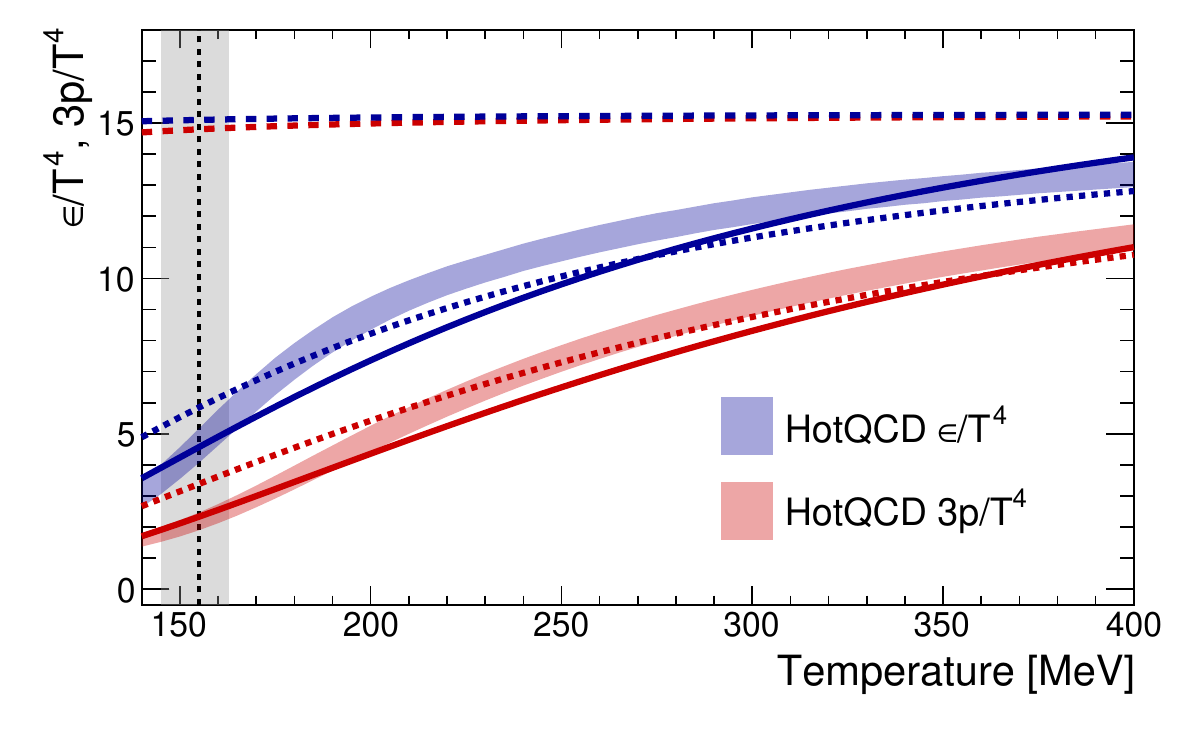} &
		\includegraphics[width=0.5\textwidth]{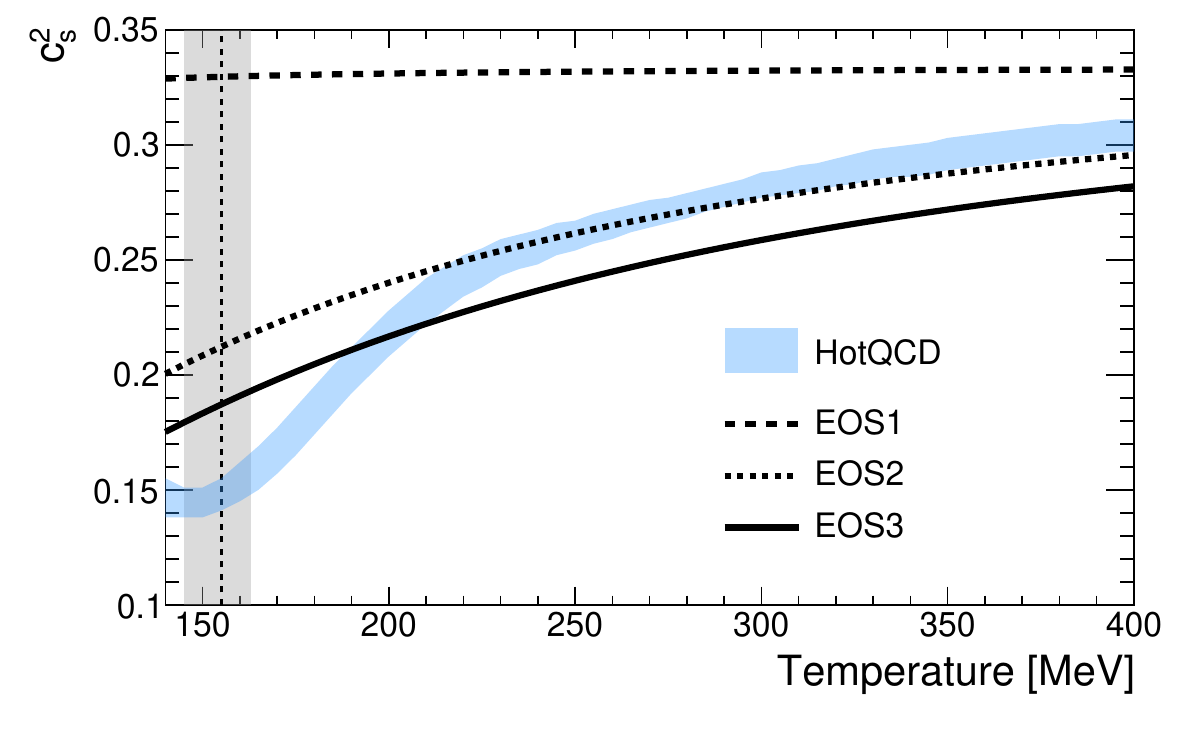}
	\end{tabular}
\caption{The values of $\epsilon/T^4$ and $3P/T^4$ (left) and the speed of sound squared (right) as functions of temperature. Three different model EOSs, obtained by varying the mass and degeneracy factor as listed in Table~\ref{tab:EOS}, are compared with the results from the HotQCD collaboration~\cite{HotQCD:2014kol}.
	The vertical gray band indicates the crossover temperature range $145<T<163\,$MeV, and the vertical gray dashed line indicates the switching temperature $T_\text{sw} = 155\,$MeV used in the event-by-event simulations.
}
	\label{fig:EOS_cs2}
\end{figure*}

Using \texttt{CoMBolt-RTA}, the effect of a nonconformal EOS on the dynamics of a system with a transverse spatial profile toward equilibrium is discussed. An event-by-event study is also performed using the hybrid version of the model. In the hybrid version, the code starts with an initial-state model to generate the initial energy density (in the present work, \trento{}~\cite{Moreland:2014oya}). After evolving the partonic state using \texttt{CoMBolt-RTA}, the description is switched to a hadron gas and the evolution continues with \texttt{UrQMD}~\cite{Bass:1998ca,Bleicher:1999xi}. It is shown that the model can describe rather well the recent OO collision data for elliptic and triangular flow, the centrality dependence of charged multiplicity, the $p_T$-spectrum, and $\la p_T\ra$ measured by ALICE~\cite{ALICE:2025luc} and ATLAS~\cite{ATLAS:2026zgq}. It is also discussed whether OO collisions are mostly in the fluid-like or particle-like regime. Using the same tuning, a prediction for potential He--He collisions at the LHC is presented.

The paper is organized as follows: In section~\ref{sec:BasicTheory}, the theoretical background of the model is discussed. The numerical method of solving the Boltzmann equation is presented in section~\ref{sec:numeric}. The evolution of a single event and a detailed study of the system's dynamics toward equilibrium are discussed in section~\ref{sec:equilibNonequilib}. In section~\ref{sec:EbyEOO}, an event-by-event simulation is performed using a hybrid version of the code to explain recent OO collision measurements at the LHC and to study the potential of even lighter ion collisions, ${}^3$He--${}^3$He and ${}^4$He--${}^4$He, in the future. The opacity parameter and its scaling for different species are presented in section~\ref{sec:OpacityScaling}. A summary and outlook are presented in section~\ref{sec:conclusion}.

\section{Massive quasiparticle dynamics from kinetic theory}\label{sec:BasicTheory}

In this section, the theoretical background of the equation to be solved is presented. The connection to the more familiar hydrodynamic quantities in equilibrium and close to equilibrium is discussed.

\subsection{(2+3)D Boltzmann equation in the relaxation time approximation}

The system has two spatial dimensions (boost invariance) and three momentum dimensions (massive quasiparticle momentum). In the present study, the quasiparticle mass is assumed to be constant. Here, the coordinate system used in Refs.~\cite{Florkowski:2013lya,Jaiswal:2021uvv} is adopted. The Boltzmann equation with Bjorken symmetry for both massless and massive particles has been studied in these references. For position space, Milne coordinates are employed, $t = \tau \cosh \eta$ and $z = \tau \sinh \eta$, where $\eta$ is the space-time rapidity. For momentum space, boost-invariant variables are used: $w = t p^z - z p^t$ and $v = t p^t - z p^z$. The distribution function is expressed as $f(\tau,x_\perp,p_T,\phi_p,w)$, where $x_\perp=(x,y)$. Using these coordinates, the Boltzmann equation can be written as
\begin{equation}\label{eq:boos-invariant-BoltzEq}
	\partial_\tau f + \frac{p_T}{p^\tau} \left( \cos\phi_p \,\partial_x f + \sin\phi_p \,\partial_y f \right)
	= \frac{C[f]}{p^\tau},
\end{equation}
where $p^\tau = \sqrt{m^2 + p_T^2 + (w/\tau)^2}$ and
\begin{equation}\label{eq:Cf}
	C[f] = -\frac{p^\mu u_\mu}{\tau_{\text{relax}}(\tau,x_\perp)}\left( f - f_{\text{eq}} \right).
\end{equation}

The energy-momentum tensor is obtained as
\begin{equation}\label{eq:Tmunu}
	T^{\mu\nu}(x_\perp) = \int_p p^\mu p^\nu f(x_\perp, p_T, \phi_p, w),
\end{equation}
where the integration measure is defined as
\begin{equation}
	\int_p \equiv \frac{1}{(2\pi)^3}\int \frac{p_T \, dp_T \, d\phi_p \, dw/\tau}{p^\tau}.
\end{equation}
The local energy density $\epsilon$ and the collective four-velocity $u^\mu$ are obtained via the Landau matching condition, $T^{\mu\nu} u_\nu = \epsilon u^\mu$.

The energy-momentum tensor can also be parameterized as
\begin{equation}
	T^{\mu\nu}
	= \epsilon u^\mu u^\nu
	- P_{\text{iso}}\,\Delta^{\mu\nu}
	+ \pi^{\mu\nu},
\end{equation}
where $\Delta^{\mu\nu} = g^{\mu\nu} -u^\mu u^\nu$ and $P_{\text{iso}} = -\frac{1}{3} \Delta_{\mu\nu} T^{\mu\nu}$. Introducing the pressure of the equilibrated system,
\begin{equation}
	P_{\text{eq}} = \frac{1}{3} \int_p |\boldsymbol{p}|^2 f_{\text{eq}}(p),
\end{equation}
where $|\boldsymbol{p}|$ is the momentum magnitude in the local rest frame, one obtains the bulk viscous pressure and shear stress tensor as
\begin{subequations}
	\begin{align}
		\Pi
		&= P_{\text{iso}} - P_{\text{eq}},\\
		\pi^{\mu\nu}
		&=  T^{\mu\nu}
		- \epsilon u^\mu u^\nu
		- \left(\Pi + P_\text{eq}\right)\,\Delta^{\mu\nu}.
	\end{align}
\end{subequations}
These quantities are particularly important for testing how far the system is from equilibrium. Furthermore, in the region where the description of the medium is switched from partons to hadrons, these quantities play an important role in constructing hadron distributions that properly match the partonic stage.

\begin{figure*}[t!]
	\centering
	\begin{tabular}{cc}
		\includegraphics[width=0.5\textwidth]{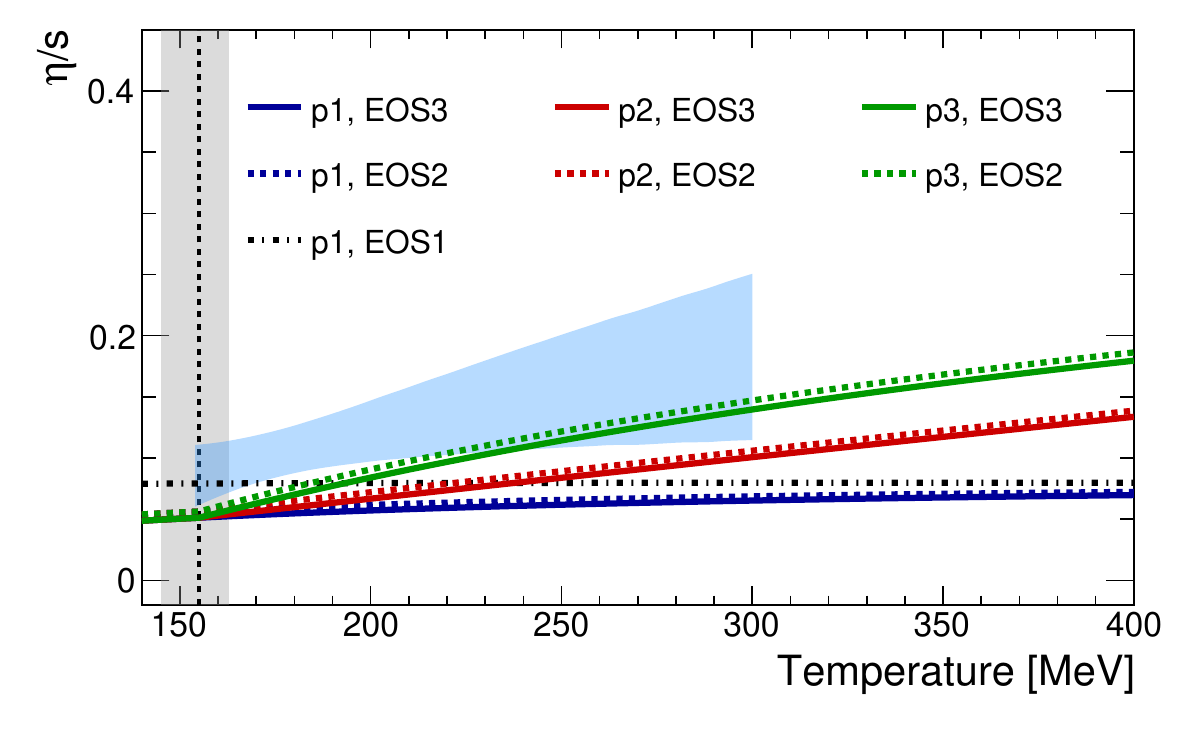} &
		\includegraphics[width=0.5\textwidth]{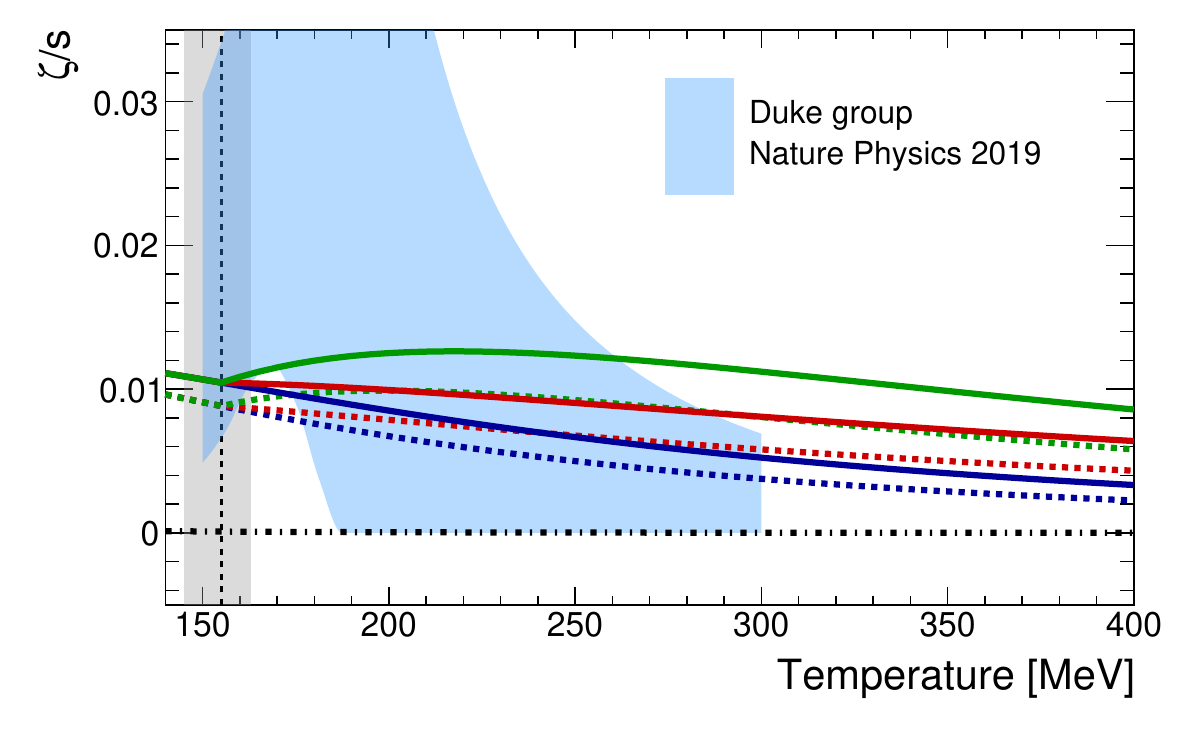}
	\end{tabular}
\caption{The specific shear viscosity $\eta/s$ (left) and bulk viscosity $\zeta/s$ (right) as functions of temperature. The model results are compared with those inferred from a Bayesian analysis using a hydrodynamics-based model in Ref.~\cite{Bernhard2019}.}
	\label{fig:shear_bulk}
\end{figure*}

\subsection{Initial state}\label{sec:IS}

The initial state is a (2+3)D Boltzmann distribution. Various scenarios can be followed to construct a full initial state. For instance, in the massless version of the model, which is available in the GitHub repository~\cite{CoMBoltGithub}, there is an option to construct the distribution using the AMPT event generator~\cite{Lin:2004en}, with a more detailed and fluctuating initial state. For the purpose of the present study, however, a minimalist approach is considered, in which one can construct an initial state from the spatial energy density by adding only one extra free parameter related to the longitudinal momentum anisotropy.

For the momentum part of the initial state, the Romatschke--Strickland (RS) distribution~\cite{Romatschke:2003ms,Jaiswal:2021uvv} is used:
\begin{equation}\label{eq:RomatStri}
	\begin{aligned}
		&f_{\text{RS}}(p_T, \phi_p, w)
		= \\ &\left[
		g_B
		\exp\!\left(
		\frac{\sqrt{m^2+p_T^2 + (1+\xi_0)(w/\tau_0)^2}}{\Lambda_0(x,y)}
		\right)
		- 1
		\right]^{-1},
	\end{aligned}
\end{equation}
where $\Lambda_0$ is a dimensionful scale and $\xi_0$ controls the initial momentum anisotropy. The initial distribution function is then given by
\begin{equation}
	f_{\text{ini}}(x,y,p_T, \phi_p, w) = f_{\text{RS}}(p_T, \phi_p, w;\Lambda_0(x,y)),
\end{equation}
where $\Lambda_0(x,y)$ is obtained by solving the following equation at each point in the transverse plane:
\begin{equation}
	\epsilon_\text{RS}(x,y,\Lambda_0(x,y)) = \epsilon_\text{ini}(x,y).
\end{equation}
Here, $\epsilon_\text{RS}$ is the energy density obtained from the RS distribution, and $\epsilon_\text{ini}(x,y)$ is the target energy density from the spatial initial-state model, namely \trento{}, MC-Glauber, or an analytical Gaussian.

\subsection{Medium in and near equilibrium}

\subsubsection{Medium in equilibrium}

The presence of a mass scale in the present study allows one to obtain an EOS closer to that obtained from lattice QCD. It is clear that, within the current approach of considering the dynamics of on-shell quasiparticles with a constant mass, the full temperature dependence of the QCD EOS cannot be exactly reproduced, particularly around the crossover region. However, by tuning the two constants $g_B$ and $m$, corresponding to the degeneracy factor and quasiparticle mass, respectively, one can approximately reproduce the lattice EOS.

The ratios $\epsilon/T^4$ and $3P/T^4$ are shown in Fig.~\ref{fig:EOS_cs2}~(left), and the speed of sound squared, $c_s^2$, is shown in Fig.~\ref{fig:EOS_cs2}~(right). A bosonic equilibrium distribution is assumed for the system,
\begin{equation}\label{eq:equilibrium}
	f_{\text{eq}}(p)
	=
	f_{\text{BE}}(p^\mu u_\mu/T),
	\qquad
	f_{\text{BE}}(x)
	=
	g_B\left[e^{x}-1\right]^{-1}.
\end{equation}
The values of the mass $m$ and the degeneracy factor $g_B$ are tuned to approximately reproduce the HotQCD lattice equation of state~\cite{HotQCD:2014kol}. It is important to note that the HotQCD EOS includes both quark and gluon degrees of freedom, while here only a single type of bosonic quasiparticle is considered. A more realistic treatment involving multiple quasiparticle species is left for future studies.

\begin{table}[t!]
	\centering
	\begin{tabular*}{\columnwidth}{@{\extracolsep{\fill}}ccc}
		\hline
		EOS num. & $g_B$ & $m$ [GeV] \\
		\hline
		1 & 46.5 & 0.05 \\
		2 & 46.5 & 0.50 \\
		3 & 55.0 & 0.62 \\
		\hline
	\end{tabular*}
\caption{Three different choices of the quasiparticle mass and degeneracy factor corresponding to three different EOS.}
	\label{tab:EOS}
\end{table}

Three different choices for $g_B$ and $m$ are listed in Table~\ref{tab:EOS}, and the corresponding EOSs are shown in Fig.~\ref{fig:EOS_cs2}. In particular, the EOS3 parameterization shows good agreement with the lattice-QCD results close to the parton--hadron switching temperature. The speed of sound shows a more complicated behavior. Compared with EOS3, the lattice-QCD EOS has a smaller speed of sound squared at low temperatures and a slightly larger value at higher temperatures.  Nevertheless, among the three parameterizations, EOS3 provides the closest approximation to the lattice-QCD EOS. In the small-mass case, EOS1, the ratios $\epsilon/T^4$ and $P/T^4$ have very weak temperature dependence, $\epsilon \approx 3P$, and the speed of sound squared is close to $1/3$, as expected for a system close to conformal symmetry.

\subsubsection{Medium near equilibrium}

In the RTA, the relaxation time $\tau_{\text{relax}}$ controls the relaxation toward equilibrium (see Eq.~\eqref{eq:Cf}). The relaxation time is parameterized as
\begin{equation}\label{eq:TauRelax}
	\tau_{\text{relax}} = \frac{5\hat{H}(T_\text{eff})}{T_\text{eff}}.
\end{equation}
In the current version of the model, $\hat{H}(T_\text{eff})$ is the quantity that controls the dissipative effects.

In the near-equilibrium limit, this quantity can be related to the more familiar transport coefficients through the following relations:
\begin{subequations}\label{eq:transport}
	\begin{align}
		\frac{\eta}{s}(T) &=
		\frac{\tau_\text{relax}(T)}{\epsilon+P_\text{eq}}
		\frac{1}{15}
		\int_p \frac{1}{p^\tau} |\boldsymbol{p}|^4 \left( -f'_\text{BE}(p/T) \right),
		\label{eq:etaOs} \\
		\frac{\zeta}{s}(T) &=
		\frac{5}{3} \frac{\eta}{s}(T)
		- \tau_\text{relax}(T) T c_s^2(T),
		\label{eq:zetaOs}
	\end{align}
\end{subequations}
These relations are slightly modified compared with those in Ref.~\cite{Jaiswal:2014isa}, where the Maxwell--Boltzmann equilibrium distribution $f_0=e^{-E/T}$ is used instead of the derivative of the Bose--Einstein distribution, $-f'_\text{BE}(E/T)$.\footnote{Eqs.~\eqref{eq:transport} are obtained by slightly modifying the calculations reported in Ref.~\cite{Jaiswal:2014isa}, using $\delta f=f-f_0$, where $f_0\equiv f_\text{BE}$ is now the Bose--Einstein distribution. To this end, one may define an integral $J_{nq}^{(r)}$, similar to $I_{nq}^{(r)}$ in Ref.~\cite{Jaiswal:2014isa}, but with the distribution function in the integrand replaced by $-f'_\text{BE}$.} Here, $c_s^2=\partial P_\text{eq}/\partial\epsilon$ is the speed of sound squared in the medium. For massless quasiparticles with the Maxwell--Boltzmann equilibrium distribution $e^{-E/T}$, one finds $\eta/s=\tau_\text{relax}T/5$ and $\zeta/s=0$.

\begin{figure}[t!]
	\centering
	\includegraphics[width=0.5\textwidth]{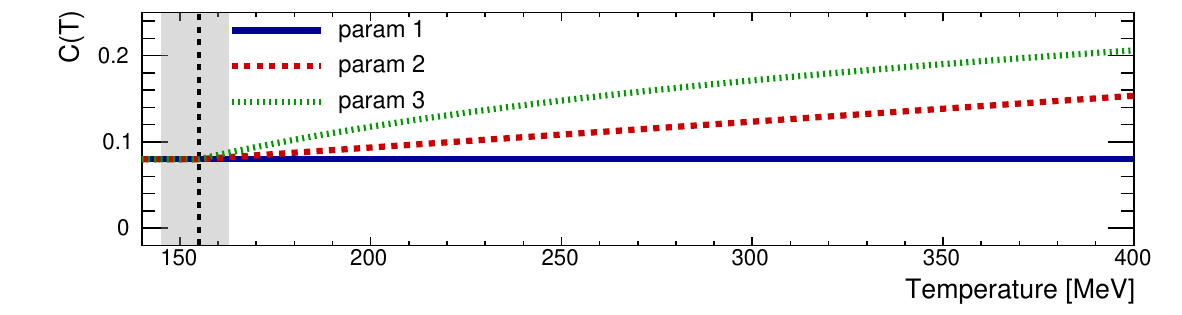}
\caption{Temperature dependence of $\hat{H}(T_\text{eff})$ for the three different parameterizations listed in Table~\ref{tab:HTeff}, based on Eq.~\eqref{eq:HTeff}.}
	\label{fig:Hparam}
\end{figure}

Since in the massless limit $\hat{H}$ behaves as the specific shear viscosity, the following parameterization is assumed:
\begin{equation}\label{eq:HTeff}
	\begin{aligned}
		&\hat{H}(T_\text{eff}) = \\
		&\begin{cases}
			\hat{H}_\text{min}+\hat{H}_\text{slop}(T_\text{eff}-T_c)\left(\frac{T_\text{eff}}{T_c}\right)^{\hat{H}_\text{crv}} & T_\text{eff} \geq T_c \\
			\hat{H}_\text{min} & T_\text{eff}< T_c
		\end{cases},
	\end{aligned}
\end{equation}
inspired by previous Bayesian studies in Ref.~\cite{Bernhard2019}. Three different parameterizations are listed in Table~\ref{tab:HTeff} and shown in Fig.~\ref{fig:Hparam}.

\begin{table}[t!]
	\centering
	\begin{tabular*}{\columnwidth}{@{\extracolsep{\fill}}ccccc}
		\hline
		\rule{0pt}{2.7ex}
		$\hat{H}$ param. & $T_c\,$[GeV] & $\hat{H}_\text{min}$ &
		$\hat{H}_\text{slop}\,$[GeV$^{-1}$] & $\hat{H}_\text{crv}$ \\
		\hline
		1 & 0.155 & 0.08 & 0.0 &  0.0 \\
		2 & 0.155 & 0.08 & 0.3 &  0.0 \\
		3 & 0.155 & 0.08 & 1.0 & -0.7 \\
		\hline
	\end{tabular*}
	\caption{Three different parameterizations of $\hat{H}(T_\text{eff})$ used in the present study.}
	\label{tab:HTeff}
\end{table}

The values of the specific shear and bulk viscosities are compared with those reported in Ref.~\cite{Bernhard2019} in Fig.~\ref{fig:shear_bulk} for three different parameterizations of $\hat{H}(T_\text{eff})$, as listed in Table~\ref{tab:HTeff}. The parameters are manually tuned to describe the recent charged-multiplicity and anisotropic-flow measurements in OO collisions by the ALICE collaboration (see Sec.~\ref{sec:EbyEOO} for details).

In Fig.~\ref{fig:shear_bulk}~(left), one sees that for the small-mass case, EOS1, one recovers $\hat{H}\approx \eta/s$ and $\zeta/s \approx 0$, as expected from conformal symmetry. For non-vanishing values of the mass, however, the specific bulk viscosity acquires sizable values comparable to those inferred from the Bayesian analysis. By modifying $\hat{H}(T_\text{eff})$, one can reproduce either a desired $\eta/s$ or $\zeta/s$. Once one of these transport coefficients is fixed, the other is automatically determined.

\section{Numerical method}\label{sec:numeric}

\begin{figure}[t!]
	\centering
	\includegraphics[width=0.5\textwidth]{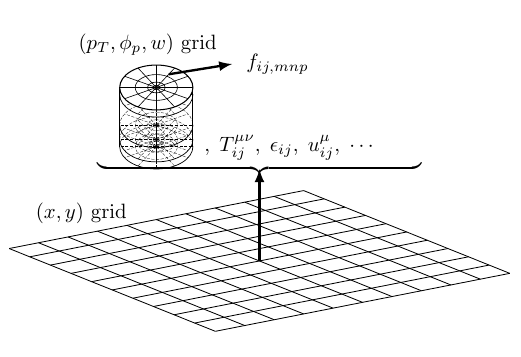}
\caption{Schematic representation of the (2+3)D grid used in the present study.}
	\label{fig:grid_schem}
\end{figure}

The method of characteristics is used to solve the equations numerically, as discussed in Refs.~\cite{Kurkela:2019kip,Taghavi:2025xhl,Taghavi:2025ddm}. In particular, a five-step algorithm is described in Ref.~\cite{Taghavi:2025xhl}, which is also followed here. The main difference is in the momentum grid. For massless quasiparticles, the angular information of the momentum encoded in $(v_z,\phi_p)$ was sufficient, while for the massive case, one needs to keep all the momentum information. A schematic representation of the (2+3)D grid is shown in Fig.~\ref{fig:grid_schem}.

Given that there are no momentum derivatives in Eq.~\eqref{eq:boos-invariant-BoltzEq}, the free-streaming coordinates $(x_0,y_0)$ are related to $(x,y)$ via
\begin{equation}\label{eq:FSCoord}
	\begin{aligned}
		&x = x_0 + V(\tau,\tau_0,p_T,w) \cos\phi_p\\
		&y = y_0 + V(\tau,\tau_0,p_T,w) \sin\phi_p
	\end{aligned}
\end{equation}
while the momentum components remain unchanged, $(p_T, \phi_p, w) = (p_T, \phi_p, w)_0$. In Eq.~\eqref{eq:FSCoord}, the magnitude of the shift is given by
\begin{equation}
	V(\tau,\tau_0,p_T,w) = \frac{p_T}{m_T^2} (\tau p^\tau - \tau_0 p_0^\tau),
\end{equation}
where the transverse mass is given by $m_T = \sqrt{m^2+p_T^2}$, and the initial temporal momentum component is defined as $p_0^\tau = \sqrt{m_T^2 + (w/\tau_0)^2}$.

Then, one time-step advancement is calculated from
\begin{equation}\label{eq:EulerStep}
	\begin{aligned}
		&f_{n+1}(x_\perp, p) = \\
		&\left[ f_n(x_{0,\perp}, p) + \frac{C[f_n]}{\sqrt{m_T^2 + (w/\tau_n)^2}}  (\tau_{n+1}- \tau_n) \right]_{x_{\perp,0}(x_\perp)}.
	\end{aligned}
\end{equation}
The right-hand side of Eq.~\eqref{eq:EulerStep}, inside the brackets, is calculated at the original grid points $x_{0,\perp}$, while the left-hand side is evaluated at $x_\perp$, given by Eq.~\eqref{eq:FSCoord}. To move to the next step, an interpolation is performed to map the distribution onto the original grid points $x_{\perp,0}$ (see Ref.~\cite{Taghavi:2025xhl} for more details).

Referring to Eq.~\eqref{eq:Cf}, the collective velocity, energy density, and effective temperature are needed to evaluate the collision kernel at each step. The equilibrium distribution is evaluated at the effective temperature $T_\text{eff}(x_\perp)$, which is obtained from the local energy density $\epsilon(x_\perp)$ by solving the equation $\epsilon = a_\epsilon(T_\text{eff}) T_\text{eff}^4$, where $a_\epsilon(T)$ is given by
\begin{equation}\label{eq:EOS_int}
	a_\epsilon(T) = \frac{1}{T^4} \int_p (|\boldsymbol{p}|^2 + m^2 )f_{\text{BE}}(\sqrt{|\boldsymbol{p}|^2 + m^2}/T).
\end{equation}

To perform the integration in Eq.~\eqref{eq:Tmunu}, the Gauss-Laguerre quadrature technique is used. The grid points in the $p_T$ and $w$ directions are discretized as
\begin{equation}\label{eq:grid}
	\begin{aligned}
		p_{T} &\in \{i=1,\ldots, n_{p_T}  |  x_{n_{p_T},i} / \lambda_{T}\}, \\
		w &\in \{i=1,\ldots, n_{w}  |  x_{n_{w},i} / \lambda_{w}\},\\
		\phi_p &\in \{i=1,\ldots, n_{\phi}  | 2\pi (i-1)  / n_\phi  \},
	\end{aligned}
\end{equation}
where $x_{n,i}$ are the zeros of the Laguerre polynomial, $L_n(x_i) = 0$, and $\lambda_T$ and $\lambda_w$ are two scaling constants used to increase the density of zeros. The energy-momentum tensor is written as
\begin{equation}
	T^{\mu\nu} = \frac{2\pi}{n_\phi}\sum_{i=1}^{n_{p_T}} \sum_{j=1}^{n_w} \sum_{k=1}^{n_\phi}  e^{\lambda_T p_{T,i} + \lambda_w w_j} \, G_{ijk}^{\mu\nu} \, \omega_{p_T,i} \, \omega_{w,j} ,
\end{equation}
where $\omega_{p_T,i}$ and $\omega_{w,j}$ are the Gauss-Laguerre weights given by
\begin{equation}
	\omega_{q,i} = \frac{x_{q,i}}{(n_q+1)^2 \left[ L_{n_q+1}(x_{q,i}) \right]^2},\quad q= p_T,w,
\end{equation}
and
\begin{equation}
	G_{ijk}^{\mu\nu}(x_\perp) = \frac{1}{(2\pi)^3}\frac{1}{\tau} \frac{1}{\lambda_T \lambda_w}\left[ \frac{p_T p^\mu p^\nu}{p^\tau} f(x_\perp,p)\right]_{ijk}.
\end{equation}
In the above, $[g(p)]_{ijk}$ denotes the discretization of $g(p)$ on the grid points defined in Eq.~\eqref{eq:grid}. The integral related to the EOS coefficient in Eq.~\eqref{eq:EOS_int} is evaluated using a similar Gauss-Laguerre quadrature, but in one dimension.

Varying but predetermined time steps are used, similar to the previous studies for massless quasiparticles in Refs.~\cite{Taghavi:2025xhl,Taghavi:2025ddm},
\begin{equation}
	h_n = h_\infty \left( \frac{\tau_n}{\tau_n +  h_\infty / \alpha_h} \right).
\end{equation}
The initial time steps are finer than the later ones. The growth rate and asymptotic time step are controlled by $\alpha_h$ and $h_\infty$, respectively.

Although one additional dimension is introduced in the momentum space for massive quasiparticle dynamics, the absence of momentum derivatives and the good performance of Gauss-Laguerre quadrature improve the computational efficiency of the code. Since there is no need for interpolation in momentum space, the code remains more stable even with larger time steps. For the moment, the size and density of the momentum grid in Eq.~\eqref{eq:grid}, as well as the time steps, need to be tuned based on the system size and the desired evolution time. To obtain stable results over longer evolution times, more grid points in momentum space are needed. In the present study, the stability is carefully monitored for each case, and the optimum momentum grid size and time steps are chosen to obtain results that are both fast and accurate.

\section{From far-from-equilibrium to near-equilibrium}\label{sec:equilibNonequilib}

Before studying collectivity in light-ion collisions using initial states generated via \trento{}, the effect of mass on the evolution of a system with a Gaussian initial energy density,
\begin{equation}
	\epsilon_\text{ini}(x,y) = \frac{dE^0_T/d\eta}{2\pi \tau_0\,R_0^{2}\sqrt{1-\epsilon_2^{2}}}
	e^{\left[\frac{1}{2R_0^{2}}\left(\frac{x^{2}}{\epsilon_2-1}-\frac{y^{2}}{\epsilon_2+1}\right)\right]},
\end{equation}
 is examined.
For the results shown in this section, the initial transverse energy is chosen as $dE^0_T/d\eta = 660\,$GeV, the initial size as $R_0=1.4\,$fm, and the initial ellipticity as $\epsilon_2 = 0.4$. These values are tuned to mimic the energy and size scales of central OO collisions, while a considerable initial ellipticity is chosen to study the system's evolution toward equilibrium for medium with high spatial anisotropy. For the mass and degeneracy factor, the values of EOS1 and EOS2 are chosen based on Table~\ref{tab:EOS}. The initial Boltzmann distribution is then constructed based on the procedure described in Sec.~\ref{sec:IS}, with the choice of $\xi_0 = 6.25$. For the quantity $\hat{H}$, a temperature-independent value is assumed, which is varied from 0.005 to 0.5, as indicated in the figure legends. The evolution starts at $\tau_0 = 0.01\,$fm/$c$ and continues until $\tau_{\text{max}} = 20\,$fm/$c$.

\begin{figure}[t!]
	\centering
	\includegraphics[width=1.0\linewidth]{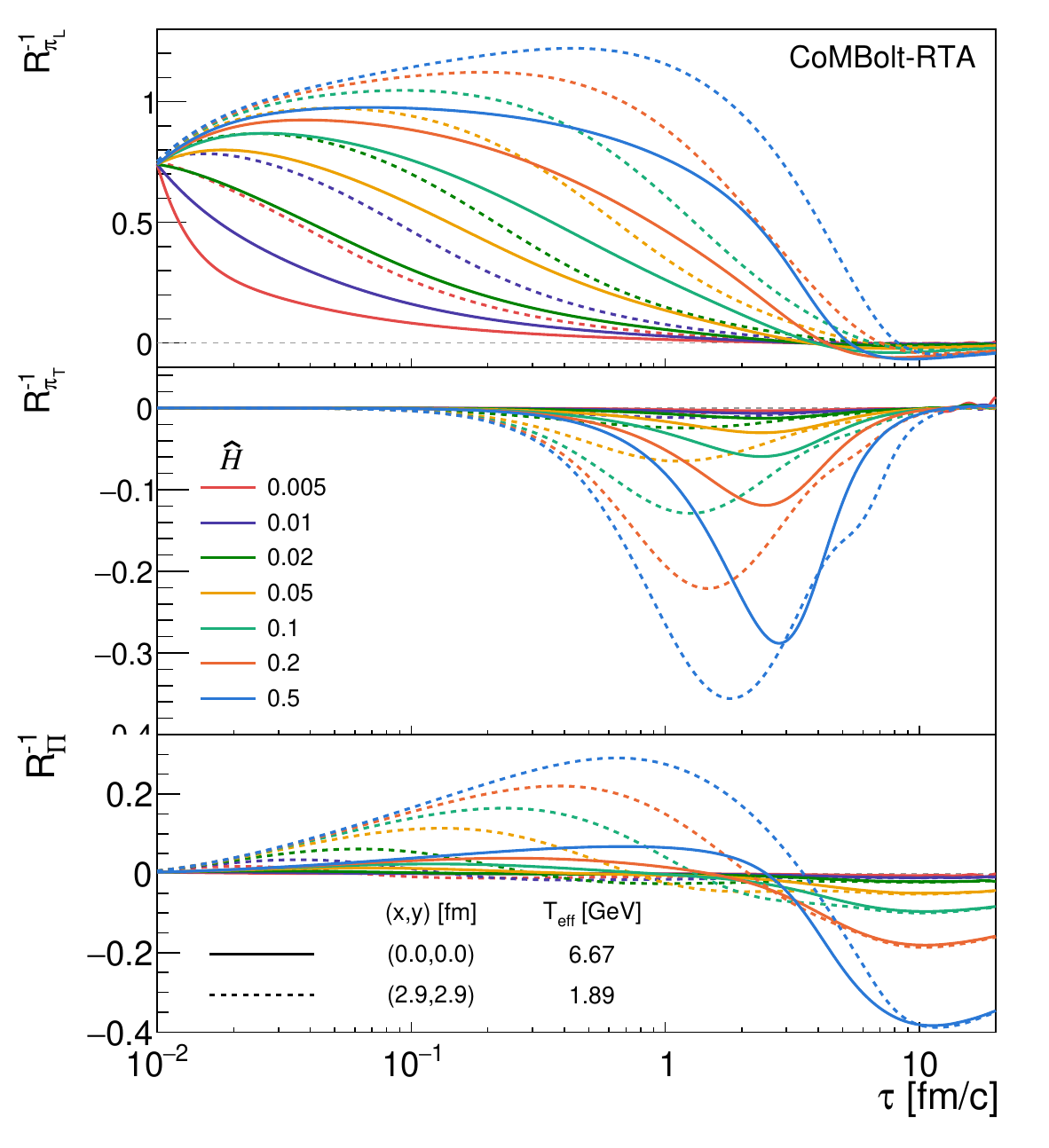}
\caption{Time evolution of the inverse Reynolds numbers for a massive system (EOS2) at two points in the transverse plane: the center (solid curves) and a point closer to the tail of the initial energy-density profile (dashed curves).}
	\label{fig:reynolds}
\end{figure}

\begin{figure*}[t]
	\centering
	\begin{tabular}{ccc}
		\includegraphics[width=0.33\textwidth]{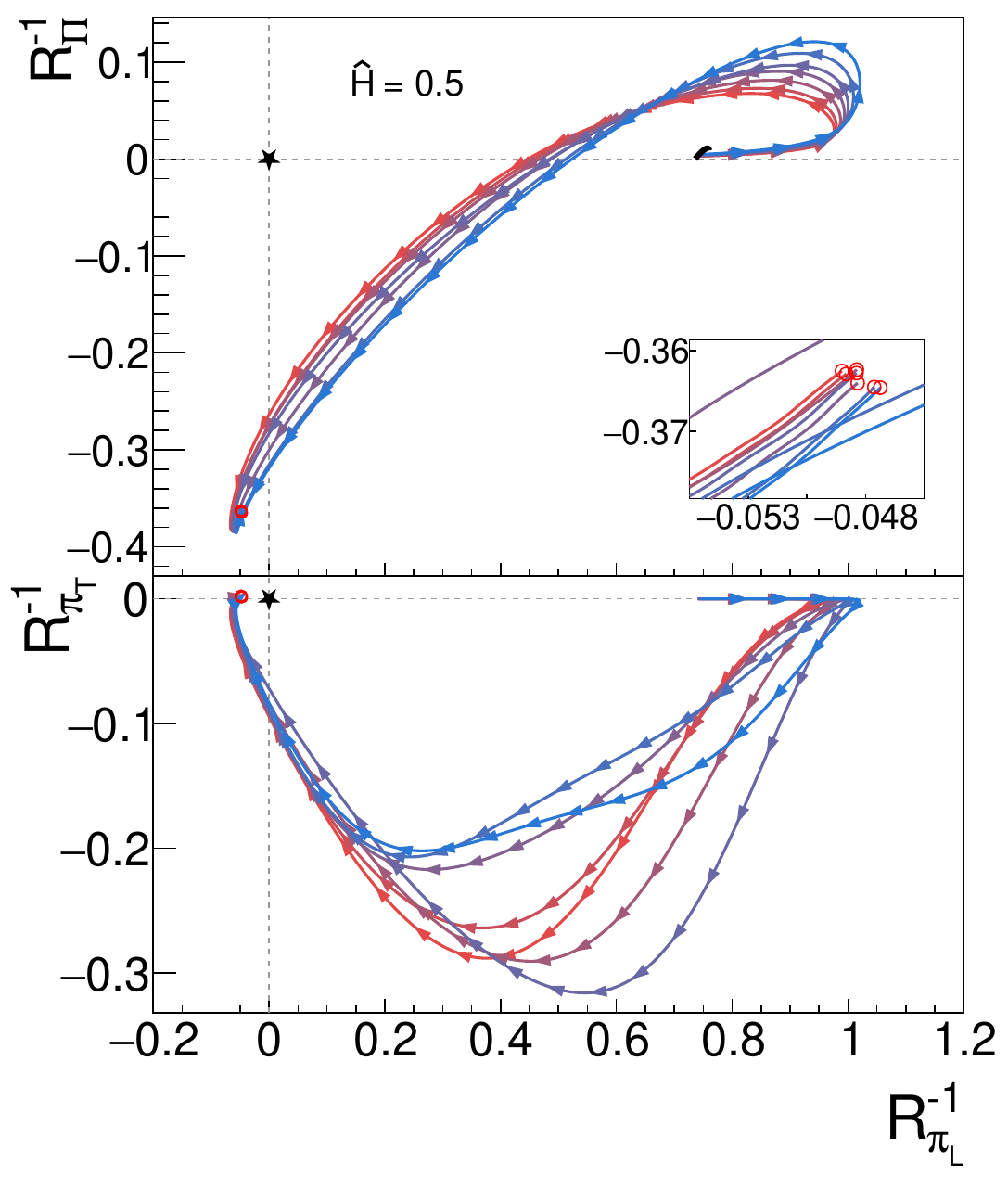} &
		\includegraphics[width=0.33\textwidth]{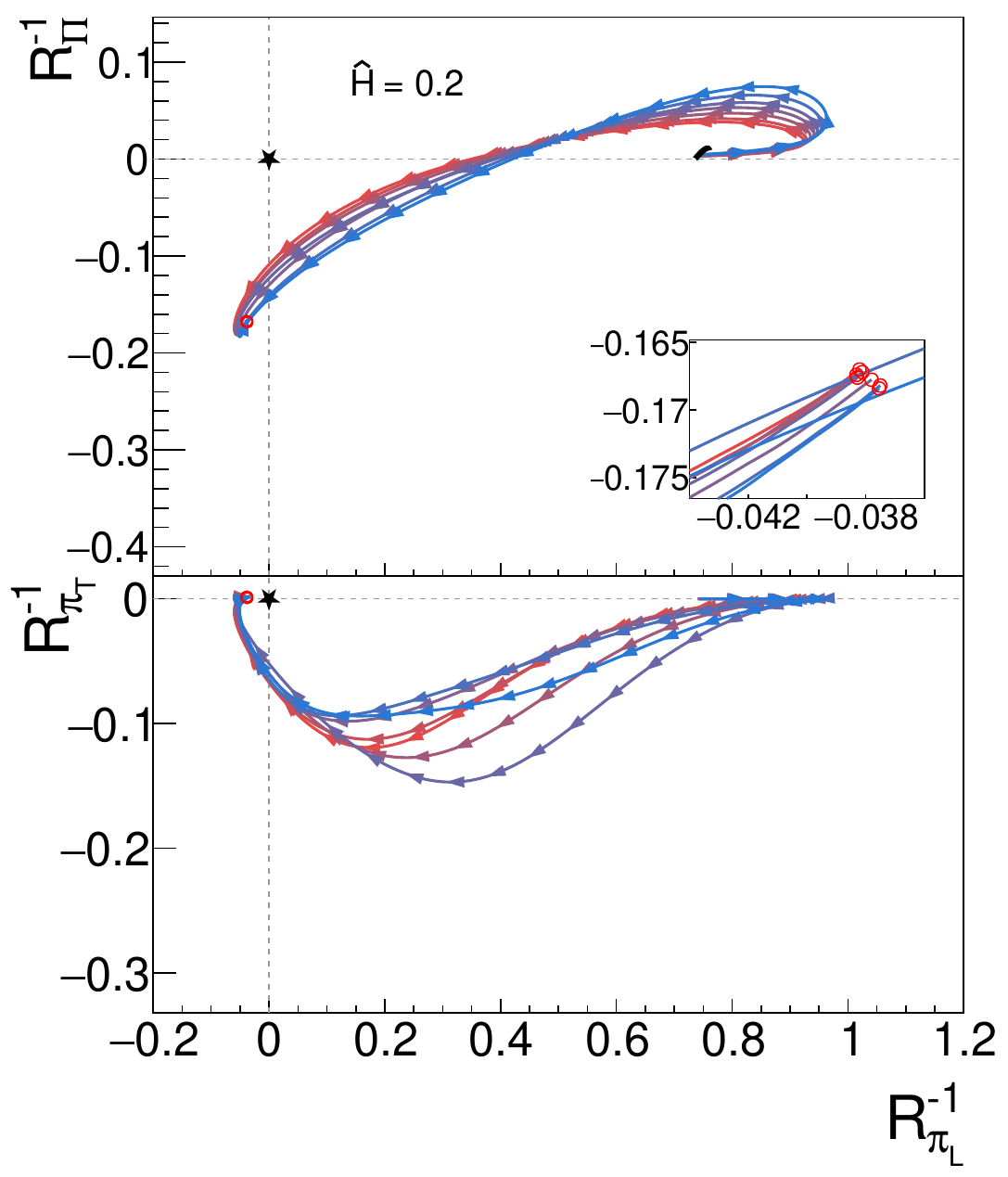} &
		\includegraphics[width=0.33\textwidth]{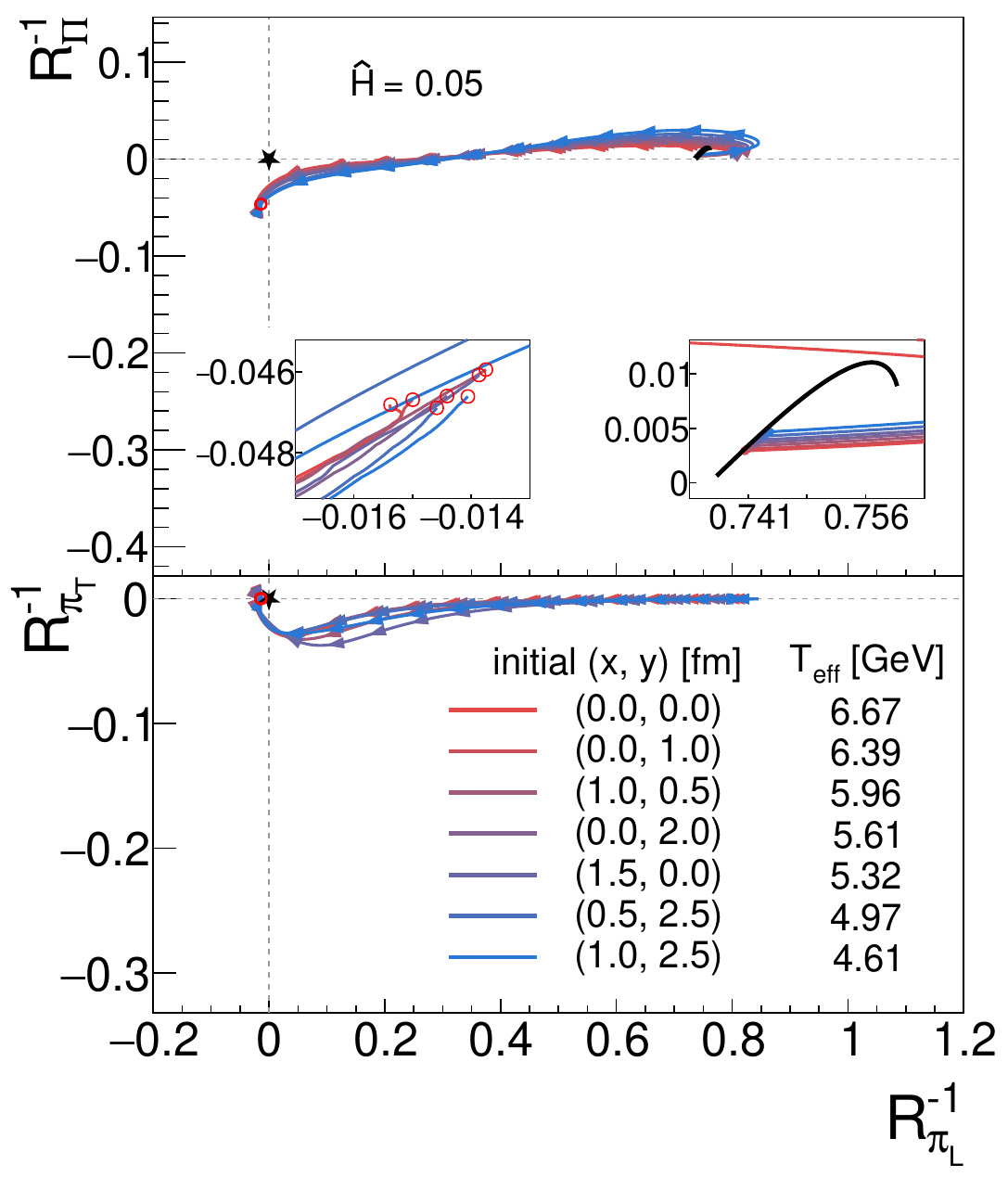}  
	\end{tabular}
\caption{Trajectories of $(R^{-1}_{\pi_L},R^{-1}_{\pi_T},R^{-1}_{\Pi})$ in anisotropy phase space for three values of $\hat{H}$. The trajectories follow longer paths toward equilibrium for systems closer to free-streaming (left), while systems closer to ideal hydrodynamics move toward equilibrium along shorter paths (right).}
	\label{fig:reynolds_phaseSpace}
\end{figure*}

In the present work, the shear and bulk inverse Reynolds numbers are defined as (see, for instance, Ref.~\cite{Denicol:2012cn})
\begin{subequations}
	\begin{align}
		R^{-1}_{\pi_L} &= \frac{P_x+P_y-2P_z}{3P_\text{eq}}, \\ 
		R^{-1}_{\pi_T} &= \frac{P_x-P_y}{\sqrt{3}P_\text{eq}}, \\
		R^{-1}_\Pi &= \frac{P_x+P_y+P_z}{3P_\text{eq}} - 1.
	\end{align}
\end{subequations}
The numerical factors are chosen such that, in the massless limit, one recovers the definition employed in previous studies~\cite{Taghavi:2025xhl,Ambrus:2022koq},
$$
\sqrt{R^{-2}_{\pi_L} + R^{-2}_{\pi_T}} = \sqrt{6\pi^{\mu\nu}\pi_{\mu\nu}/\epsilon^2}.
$$

\begin{figure*}[t]
	\centering
	\begin{tabular}{cc}
		\includegraphics[width=0.5\textwidth]{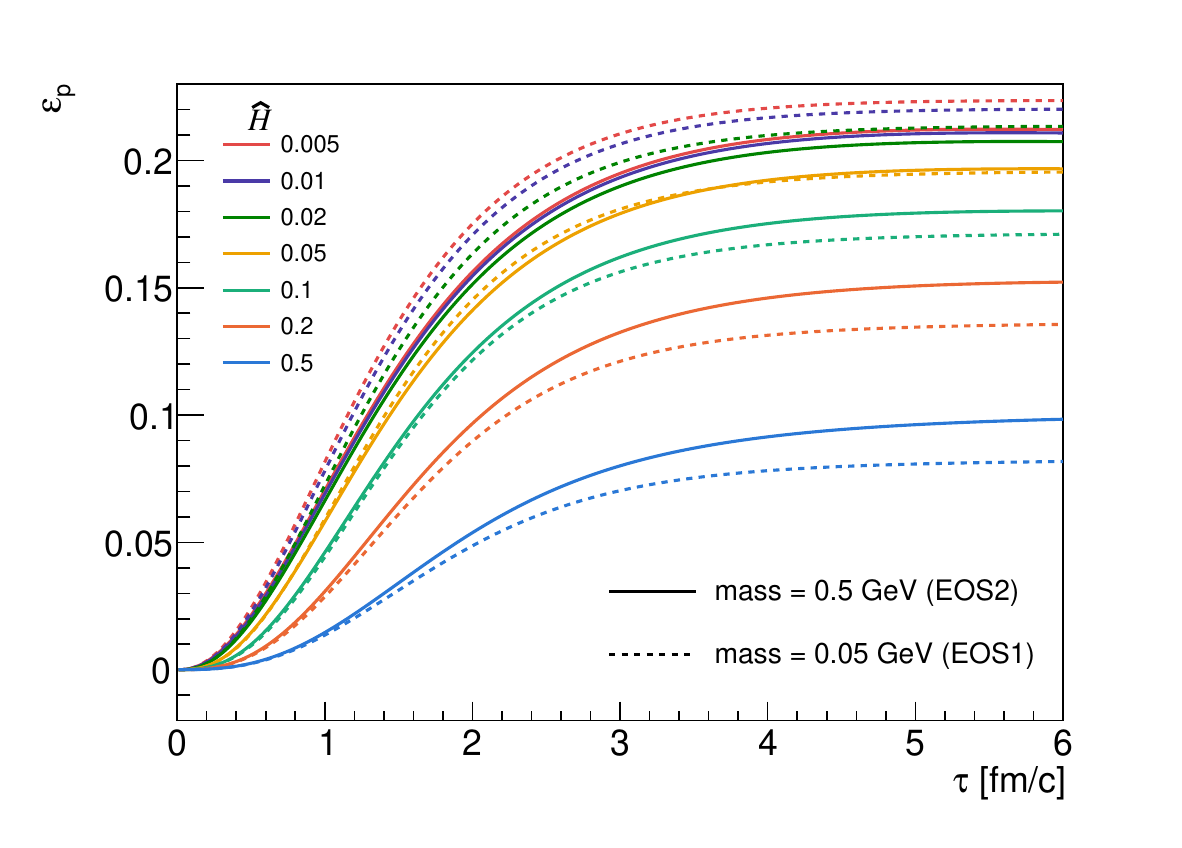} &
		\includegraphics[width=0.5\textwidth]{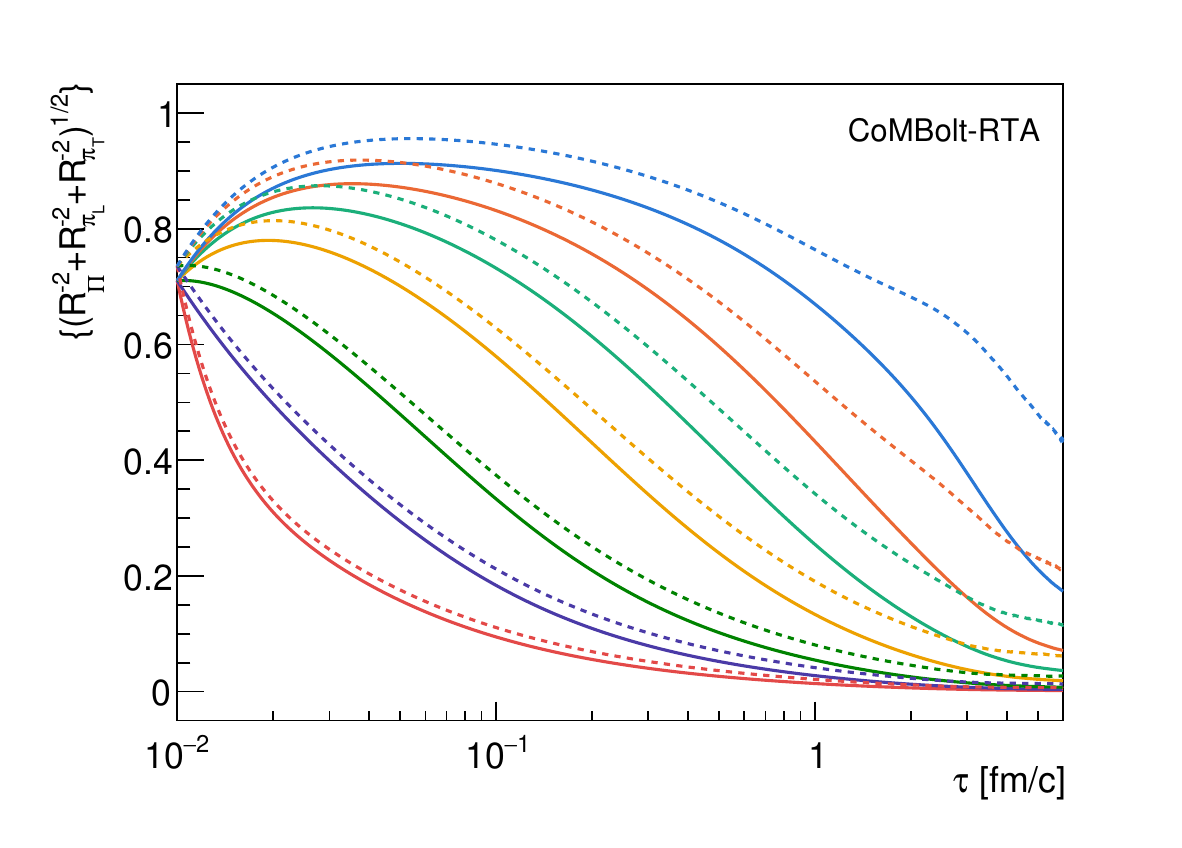}
	\end{tabular}
\caption{Energy-density-weighted quantities in the transverse plane. Left: Time evolution of the momentum anisotropy for different values of $\hat{H}$ and two quasiparticle masses, starting from identical initial conditions with a sizable initial ellipticity. The larger mass is shown by solid curves and the smaller mass by dashed curves. Right: Time evolution of the average inverse Reynolds number, representing the energy-density-weighted average distance of the point $(R^{-1}_{\pi_L},R^{-1}_{\pi_T},R^{-1}_{\Pi})$ from the origin of anisotropy phase space, which corresponds to equilibrium.}
	\label{fig:momentAnis}
\end{figure*}

The time evolution of the three inverse Reynolds numbers for a system with Gaussian initial energy density at two spatial points (the center shown by solid lines and $(2.9,2.9)$ [fm] by dashed lines) is presented in Fig.~\ref{fig:reynolds}. Smaller values of $\hat{H}$ correspond to systems closer to ideal hydrodynamics, which move toward equilibrium (zero inverse Reynolds numbers) more quickly. By construction, there is no initial transverse anisotropy in the system, $R^{-1}_{\pi_T}=0$. However, the choice of $\xi_0$ induces a sizable initial longitudinal anisotropy $R^{-1}_{\pi_L}$. One can also show that $R^{-1}_{\Pi}$ is non-vanishing initially. Due to the large anisotropy generated in systems closer to free-streaming, $\hat{H} \sim 0.5$, the small initial value of $R^{-1}_{\Pi}$ is not clearly visible in the figure.

In fact, the initial $P_\text{eq}$, $P_x$, $P_y$, and $P_z$ have been worked out for the RS distribution in Refs.~\cite{Martinez:2009ry,Florkowski:2014bba}. Based on the minimalist initial-state construction, the initial $R^{-1}_{\Pi}$ and $R^{-1}_{\pi_L}$ depend only on $T_\text{eff}$ for a fixed EOS. As a result, two separate points in the initial energy-density profile with the same $T_\text{eff}$ (initial energy density) have identical $R^{-1}_{\Pi}$ and $R^{-1}_{\pi_L}$. These initial values are shown by black curves in Fig.~\ref{fig:reynolds_phaseSpace}. The curve is more visible in the zoomed-in inset in the right panel.

Due to the transverse spatial extension and the initial transverse spatial anisotropy, the system develops a nonzero $R^{-1}_{\pi_T}$ after around $1\,\text{fm}/c$, although it is initially zero. A sizable $R^{-1}_{\Pi}$ is also generated, particularly for larger values of $\hat{H}$ and lower temperatures, where the relaxation time $\tau_\text{relax}$ is larger and the system is closer to free-streaming. In general, it is clear from the figure that the system develops more anisotropy in colder regions. Another feature of the inverse Reynolds numbers in Fig.~\ref{fig:reynolds} is that, starting from two different spatial points, their values approach each other for a fixed $\hat{H}$, as discussed in more detail in connection with Fig.~\ref{fig:reynolds_phaseSpace}. In Ref.~\cite{Jaiswal:2021uvv}, massive kinetic theory without transverse spatial dimensions is discussed, meaning that there are no separate dynamics for $R^{-1}_{\pi_L}$ and $R^{-1}_{\pi_T}$. In the massless case, there are no dynamics in the $R^{-1}_{\Pi}$ direction. The inclusion of transverse spatial dimensions in the massive quasiparticle framework leads to richer and more complex dynamics as the system evolves toward equilibrium.

The trajectories of the inverse Reynolds numbers in anisotropy phase space are shown in Fig.~\ref{fig:reynolds_phaseSpace} for seven different starting points in the transverse plane. As depicted in the figure, for a system closer to ideal hydrodynamics with $\hat{H} = 0.05$, the phase point $(R^{-1}_{\pi_L},R^{-1}_{\pi_T},R^{-1}_{\Pi})$ follows a short trajectory, while for larger values of $\hat{H}$, it traverses a longer path. As seen in the upper panels of the figure, the trajectories initially appear to move away from the equilibrium point in all cases. For systems with smaller $\hat{H}$, the deviation from $R^{-1}_{\Pi} = 0$ is smaller than for larger values of $\hat{H}$. However, looking at the insets, one observes that the system has a tendency to move toward equilibrium at late times. Due to the longitudinal expansion, however, the dynamics becomes slower and slower, preventing the system from reaching full equilibrium. This is consistent with the slow late-time evolution observed in Fig.~\ref{fig:reynolds}, which is more clearly visible on the logarithmic time axis.

\begin{table*}[ht!]
	\centering
	\begin{tabular}{cccccccc|cc}
		\hline
		Spatial (\trento) & norm & nucl. width & const. width &
		const. number & redu. thick. & fluct. & cross-section & Momentum & $\xi_0$ \\
		\hline
		& 30 & 0.45 & 0.45 & 4 & 0.0 & 2.0 & 6.8 & & 6.25 \\
		\hline
	\end{tabular}
	\caption{Spatial and momentum initial-state parameters used in the event-by-event simulations.}
	\label{tab:init_param}
\end{table*}
\begin{figure*}[ht!]
	\centering
	\begin{tabular}{cc}
		\includegraphics[width=0.5\textwidth]{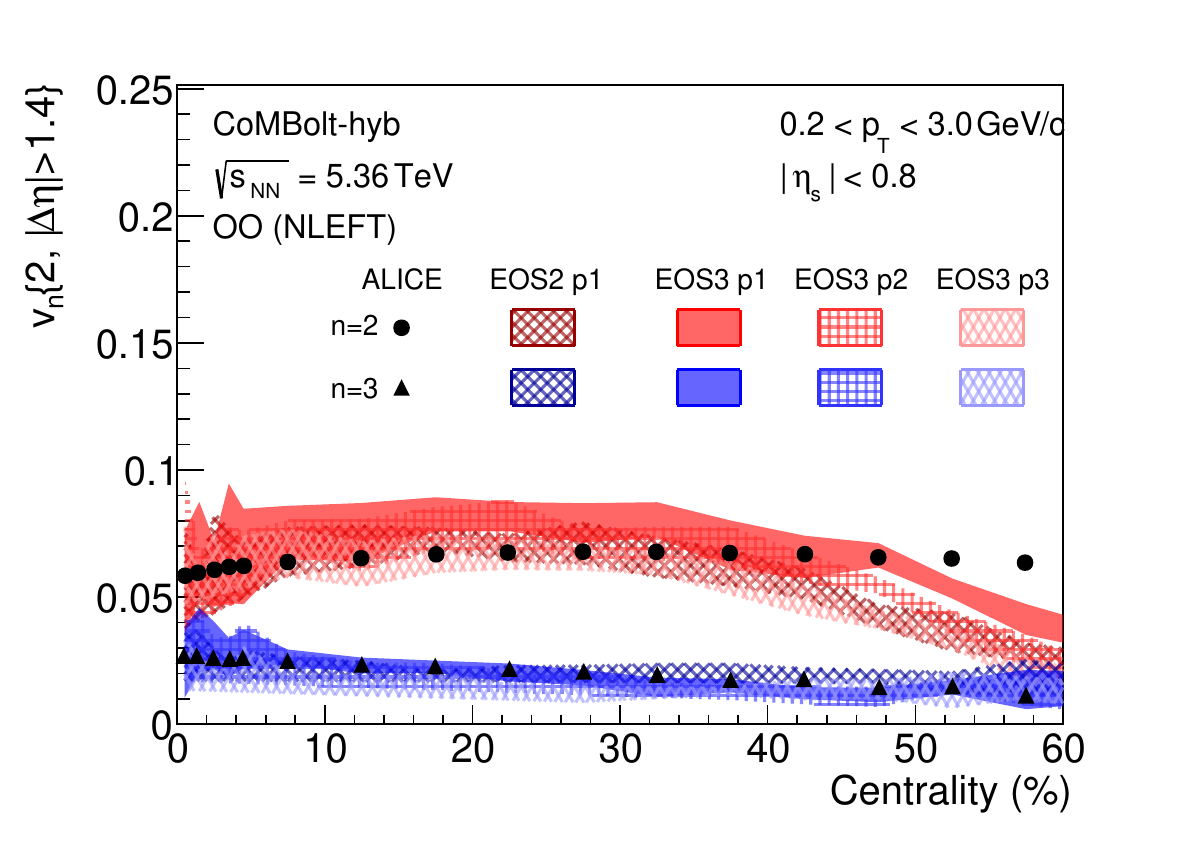} &
		\includegraphics[width=0.5\textwidth]{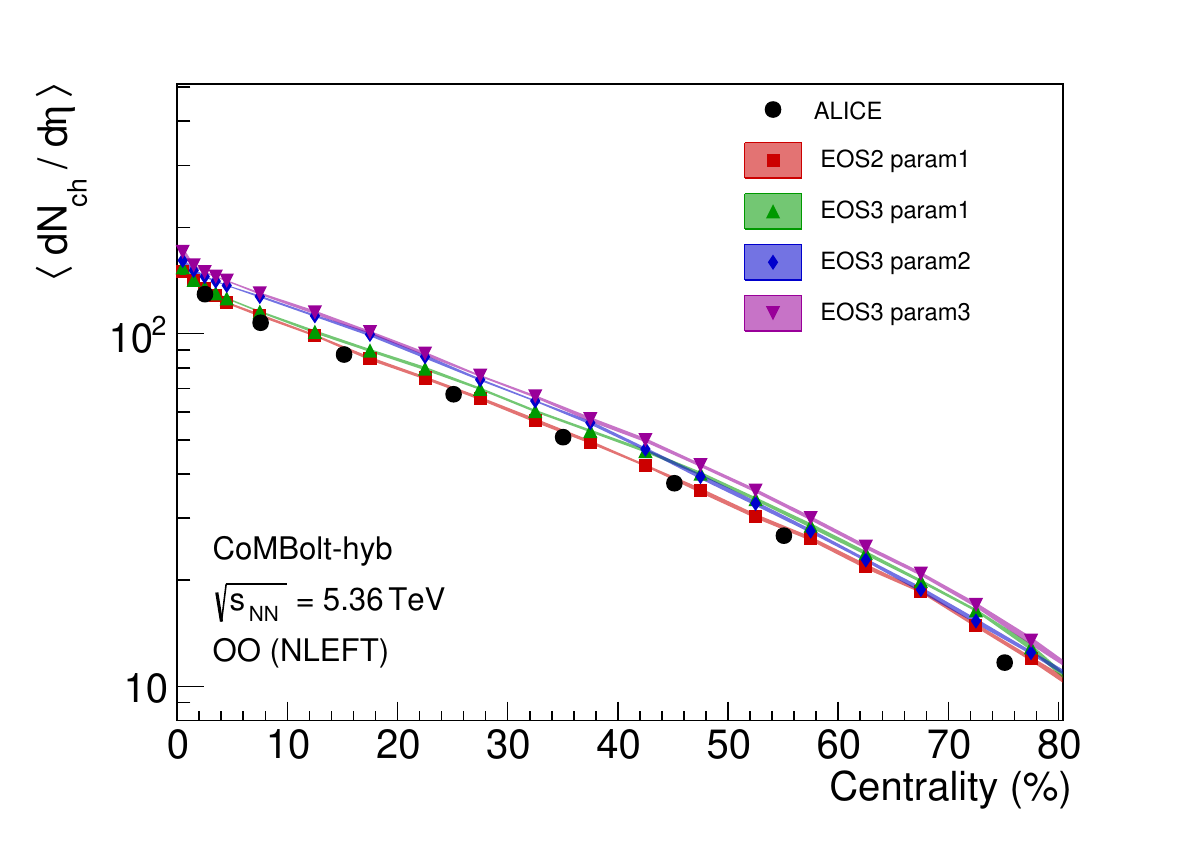}
	\end{tabular}
\caption{Anisotropic flow obtained from two-particle correlations (left) and the centrality dependence of charged multiplicity (right), calculated using \texttt{CoMBolt-hyb} and compared with ALICE measurements~\cite{ALICE:2025luc,ALICE:2025ISmult}. Different combinations of EOS and $\hat{H}$ parameterizations are presented.}
	\label{fig:v22_charged_multi}
\end{figure*}

So far, the evolution of local quantities has been discussed. One can also study the evolution of averaged quantities. In Fig.~\ref{fig:momentAnis}, the momentum anisotropy,
\begin{equation}\label{eq:momentumAnis}
	\epsilon_p(\tau) =\left|\frac{\int dx_\perp \left(T^{xx}  - T^{yy}  + 2i T_{xy} \right) }{ \int dx_\perp \left( T^{xx} +T^{yy} \right)  }\right|
\end{equation}
and the average inverse Reynolds number, $\{\sqrt{R^{-2}_{\pi_L} + R^{-2}_{\pi_T} + R^{-2}_{\Pi}} \}$, are shown for two cases, EOS1 and EOS2, and various values of $\hat{H}$. The curly brackets indicate the energy-density-weighted average,
\begin{equation}
	\{ \mathcal{O} \} = \int d^2x_\perp \mathcal{O} \epsilon(x_\perp) / \int d^2x_\perp \epsilon(x_\perp).
\end{equation}

The momentum anisotropy is a measure of the strength of the system's collective behavior in converting spatial anisotropy into momentum anisotropy. As seen in Fig.~\ref{fig:momentAnis}~(left), within the evolution time shown in the figure, a larger mass increases the value of $\epsilon_p$ for systems closer to free-streaming (larger $\hat{H}$). This behavior changes for systems closer to ideal hydrodynamics. For $\hat{H} = 0.005$, the value of $\epsilon_p$ is systematically larger for systems with smaller mass.

The average inverse Reynolds number is shown in Fig.~\ref{fig:momentAnis}~(right). This quantity can be considered as the energy-density-weighted average distance of the phase-space point $(R^{-1}_{\pi_L},R^{-1}_{\pi_T},R^{-1}_{\Pi})$ from the equilibrium point. Systems with larger mass move toward equilibrium slightly faster, as seen from the solid curves, which are below the dashed curves for all cases within the evolution time shown in the figure. More complicated late-time dynamics is observed for the small-mass case closer to free-streaming, with $\hat{H} = 0.5$, similar to what was previously observed in Ref.~\cite{Ambrus:2022koq} for massless systems. Apparently smoother dynamics is observed for larger masses.

\section{Event-by-event simulations}\label{sec:EbyEOO}

The event-by-event simulations of light-ion collisions are performed using a hybrid chain that includes \trento{} for the initial state, \texttt{CoMBolt-RTA} for the partonic evolution, and \texttt{UrQMD} as a hadronic afterburner. The switching temperature to the hadronic phase is set to $T_\text{sw}=155\,$MeV, and the evolution starts at $\tau_0 = 0.1\,$fm/$c$. The simulation setup is almost identical to that described in Ref.~\cite{Taghavi:2025ddm}, with \texttt{CoMBolt-ITA} replaced by \texttt{CoMBolt-RTA}. For each parameter set, 3000 minimum-bias events are generated for each collision system.

Given that this hybrid model could have different dynamics compared to models based on hydrodynamics, the parameter tunings of those models obtained using Bayesian analysis approaches (see, e.g., Refs.~\cite{Bernhard2019,JETSCAPE:2020mzn,Nijs:2020roc,
	Parkkila:2021tqq,Parkkila:2021yha}) are not directly applicable here. For that reason, a dedicated Bayesian analysis is required~\cite{Taghavi:BayesianOO}. Before obtaining robust results from the Bayesian analysis, a manual search over a limited number of points in the parameter space is performed here to obtain approximate values for the parameters. The target observables are $v_2\{2\}$ and the centrality dependence of $dN_\text{ch}/d\eta$ measured by the ALICE collaboration~\cite{ALICE:2025luc,ALICE:2025ISmult}. The tested parameter sets with poor agreement with the data are not shown in the present study.

Only one set of initial-state parameters is presented, as listed in Table~\ref{tab:init_param}, although more options have been tried, particularly for the overall normalization, nucleon and constituent widths, and the number of constituents. For oxygen nuclei, the nucleon configurations (NLEFT) reported in Ref.~\cite{Giacalone:2024luz} are used. The nucleon configurations for ${}^3$He and ${}^4$He are imported from PHOBOS MC-Glauber~\cite{Alver:2008aq}. The parameters related to the medium are listed in Tables~\ref{tab:EOS} and \ref{tab:HTeff}.

\subsection{ALICE and ATLAS OO collision measurements from CoMBolt-hyb}

\begin{figure*}[t]
	\centering
	\begin{tabular}{ccc}
		\includegraphics[width=0.44\textwidth]{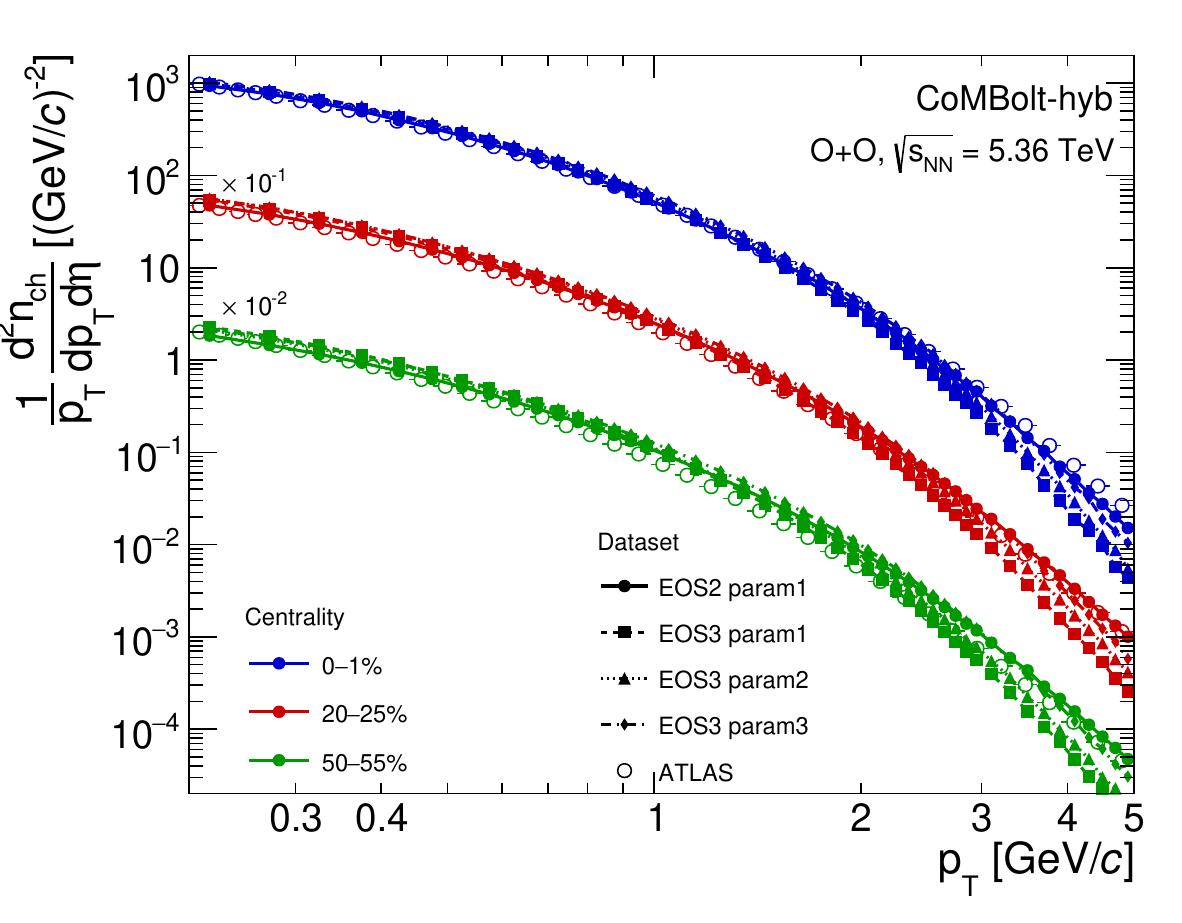} & \hspace*{0.7cm} &
		\includegraphics[width=0.5\textwidth]{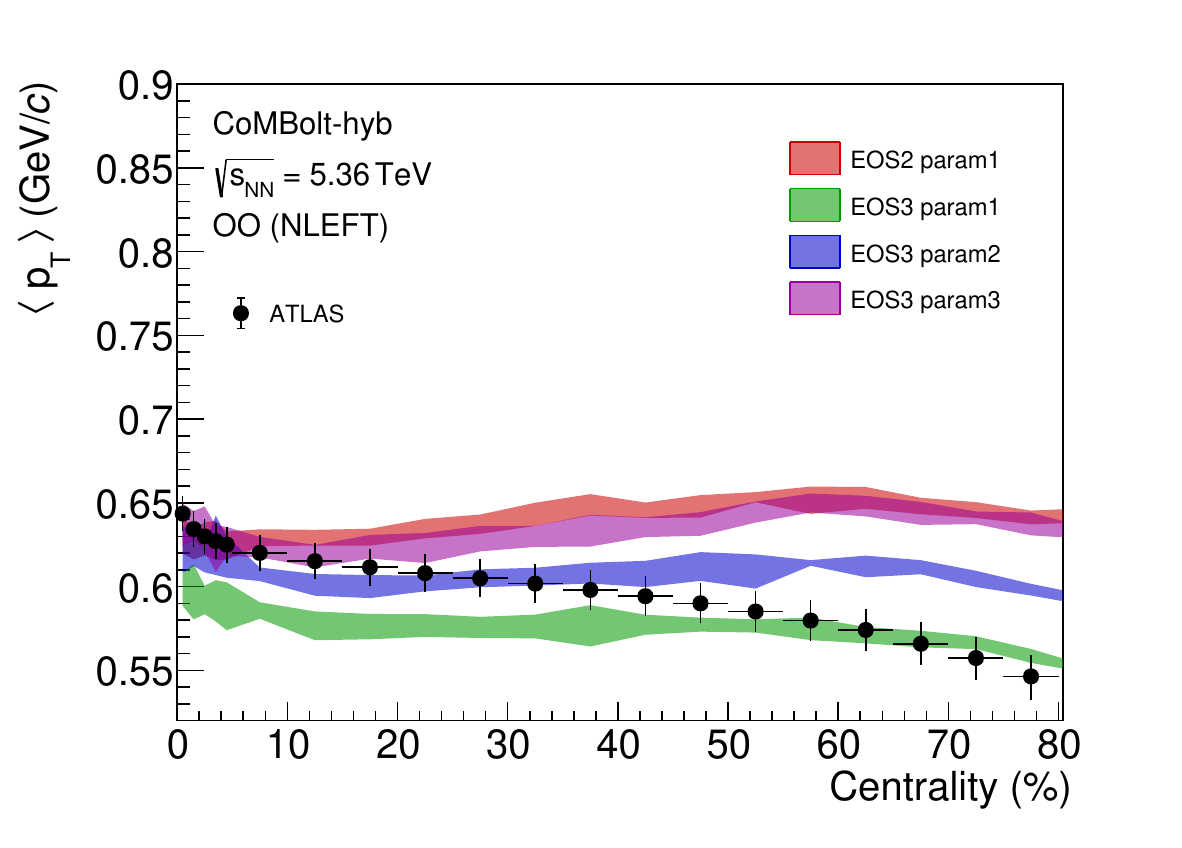}
	\end{tabular}
	\caption{Comparing the $p_T$-spectrum (left) and $\la p_T \ra$ (right) simulated by \texttt{CoMBolt-hyb} and that measured by ATLAS collaboration~\cite{ATLAS:2026zgq}. Different tuning of the model are presented same as that in Fig.~\ref{fig:v22_charged_multi}. }
	\label{fig:pt_spec_and_avPT}
\end{figure*}

Given that the final state of \texttt{CoMBolt-hyb} is obtained from \texttt{UrQMD}, a more realistic analysis of the observables with accurate kinematic cuts is applied in the present work. More details have been discussed in Ref.~\cite{Taghavi:2025ddm} and are not repeated here. The $v_2\{2\}$ and $v_3\{2\}$ predictions of \texttt{CoMBolt-hyb} are compared with ALICE measurements~\cite{ALICE:2025luc,ALICE:2025ISmult} in Fig.~\ref{fig:v22_charged_multi}~(left), and the centrality dependence of charged multiplicity is shown in Fig.~\ref{fig:v22_charged_multi}~(right).

From Fig.~\ref{fig:v22_charged_multi}~(left), one finds better agreement with the data for EOS3 and the $\hat{H}=0.08$ parameterization (param1) with no temperature dependence. As seen, none of the tested parameter tunings can describe the data over a wide range of centrality, and the agreement becomes worse at higher centralities. Referring to Fig.~\ref{fig:EOS_cs2}, EOS3 corresponds to the case where the system's energy density and pressure are closest to the HotQCD EOS at the switching temperature. Also, the speed of sound squared is underestimated at high temperatures and overestimated at lower temperatures. The first $\hat{H}$ parameterization leads to a very small temperature dependence of $\eta/s$, with a small initial value. The other EOSs and $\hat{H}$ parameterizations show worse agreement at higher centralities and also produce larger charged multiplicities, as seen in Fig.~\ref{fig:v22_charged_multi}~(right).

Potential missing physics in the modeling of either the initial state or collectivity could explain the deviations at higher centralities. It is also possible that some non-flow contributions remain in the experimental measurements and have not been fully removed.

Compared to the Bayesian analyses of models based on hydrodynamics~\cite{Bernhard2019,JETSCAPE:2020mzn,Nijs:2020roc,Parkkila:2021tqq,Parkkila:2021yha}, the manual tuning shows better agreement with the data using a smaller nucleon width (see Table~\ref{tab:init_param}) and a smaller $\eta/s$ (see Fig.~\ref{fig:shear_bulk}). Although manual tuning is an acceptable starting point, a systematic Bayesian analysis is needed to perform a more comprehensive search for the best parameter tuning~\cite{Taghavi:BayesianOO}.

After searching for a reasonable parameter tuning to explain $v_2\{2\}$ and the centrality dependence of $dN_\text{ch}/d\eta$, the model predictions for the $p_T$-spectrum and $\la p_T \ra$ are compared with the ATLAS measurements. These two observables are particularly sensitive to the EOS and could not be accurately described using the massless quasiparticle dynamics of \texttt{CoMBolt-ITA}. The $p_T$-spectrum is shown in Fig.~\ref{fig:pt_spec_and_avPT}~(left). As seen, different parameter tunings show reasonable agreement between the model and the data. Larger deviations are observed at higher $p_T$ values, where hard processes are expected to play a role, and the current model, like many other hydrodynamics-based models, is not designed to describe this region.

The average transverse momentum, $\la p_T \ra$, measured by ATLAS is compared with \texttt{CoMBolt-hyb} in Fig.~\ref{fig:pt_spec_and_avPT}~(right). As discussed in Ref.~\cite{Taghavi:2025ddm}, the previous version of \texttt{CoMBolt} systematically overestimated $\la p_T \ra$. A comparison with hydrodynamic simulations indicated that including a non-ideal EOS and bulk viscosity could reduce $\la p_T \ra$ by roughly 30\%. Including a quasiparticle mass in the present version leads to much better agreement with the data. This improvement is attributed to a more realistic description of the speed of sound in the new version. 
The connection between $\la p_T \ra$, the EOS, and the speed of sound has been extensively discussed in previous studies~\cite{Gardim:2019xjs,CMS:2024sgx}. The manual tuning does not lead to perfect agreement between the model and the data. One possible reason is the lack of an extensive search for the best parameter tuning. Another, potentially more important, reason is that the full QCD EOS has not yet been implemented in the code, which may affect the prediction for the average transverse momentum.

\subsection{Moving toward lighter-ion collisions: ${}^3$He--${}^3$He, ${}^4$He--${}^4$He, and p--Pb}

\begin{figure*}[t]
	\centering
	\begin{tabular}{cc}
		\includegraphics[width=0.5\textwidth]{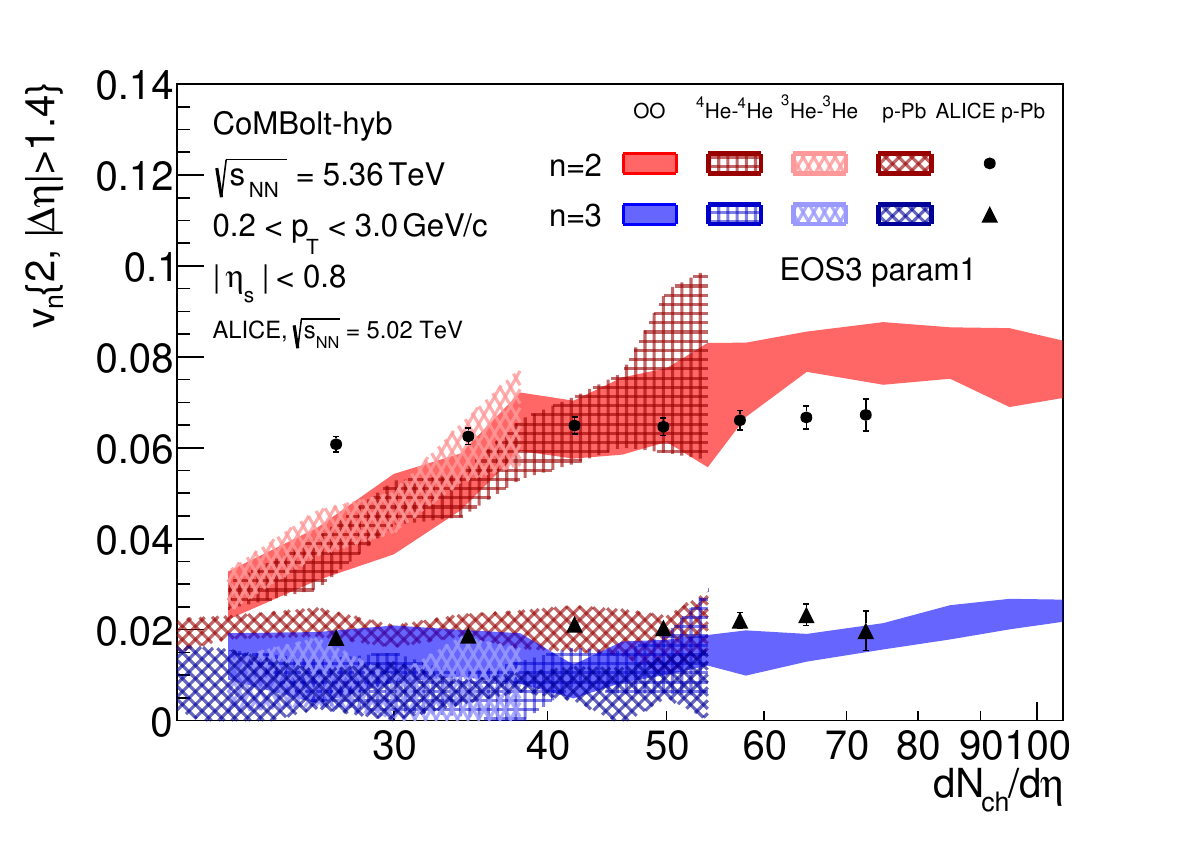}&
		\includegraphics[width=0.5\textwidth]{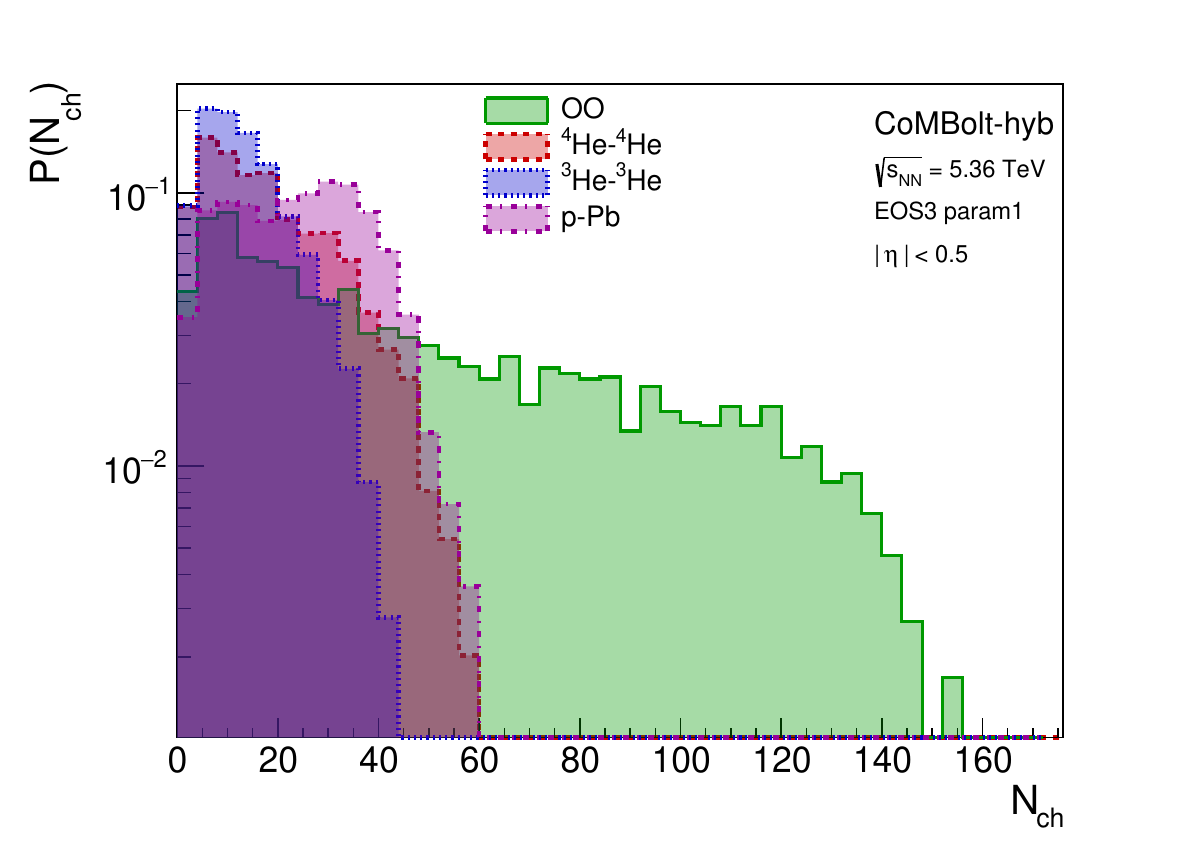} 
	\end{tabular}
\caption{Comparison of elliptic and triangular flow in lighter collision systems, ${}^3$He--${}^3$He, ${}^4$He--${}^4$He, and p--Pb, with peripheral OO collisions, calculated using \texttt{CoMBolt-hyb} with the EOS3 param1 tuning (left), and the charged-multiplicity distributions (right). The ALICE measurement of $v_2\{2\}$ in p--Pb collisions is also shown in the left panel~\cite{ALICE:2019zfl}. The center-of-mass energy of the p--Pb data is slightly different, and $dN_\text{ch}/d\eta$ is obtained using a naive scaling to match the pseudorapidity windows.}

	\label{fig:v22_system_size_scan}
\end{figure*}

Having obtained a reasonable description of OO collision measurements, the same parameter tuning can be used to study collectivity in smaller collision systems. In Ref.~\cite{Taghavi:2025ddm}, using a massless quasiparticle model, it was shown that central OO collisions are still in a regime where the medium behaves more fluid-like. There are debates about the potential of ${}^4$He--${}^4$He collisions at the LHC. This motivates testing smaller collision systems based on the tuning obtained from OO collisions. Choosing EOS3 and the first $\hat{H}$ parameterization as the best parameter tuning, the predictions for the charged-multiplicity dependence of $v_2\{2\}$ and $v_3\{2\}$ are shown in Fig.~\ref{fig:v22_system_size_scan}~(left) for ${}^3$He--${}^3$He and ${}^4$He--${}^4$He collisions.

It is seen that $v_2\{2\}$ and $v_3\{2\}$ for these two collision systems follow the same trend as OO collisions. This can be explained by the cluster decomposition of oxygen nuclei. In peripheral collisions, one expects two alpha particles in the oxygen nuclei to collide. In Fig.~\ref{fig:v22_system_size_scan}~(right), the charged-multiplicity distributions are shown, from which one can infer that the most central ${}^3$He--${}^3$He and ${}^4$He--${}^4$He collisions have similar sizes to more peripheral OO collisions.

In Fig.~\ref{fig:v22_system_size_scan}, the results for p--Pb collisions are also shown. The anisotropic flow measurements of p--Pb by ALICE are at a slightly different center-of-mass energy~\cite{ALICE:2019zfl}. Also, the reported charged-multiplicity bins have a slightly larger $\eta$ window. Assuming that the slightly different $\sqrt{s_\text{NN}}$ has a minor impact and applying a naive correction for the $\eta$ window, the results are overlaid with the predictions for OO, ${}^3$He--${}^3$He, and ${}^4$He--${}^4$He collisions. As seen in the figure, the \texttt{CoMBolt-hyb} prediction does not follow the data and is significantly smaller.

Various reasons could explain this discrepancy. The most probable issue is the inaccuracy in modeling the initial state. Here, the same \trento{} parameter tuning as for OO collisions is used, and only the target and projectile are replaced by p and Pb. The theoretical uncertainty in the initial state is more pronounced, particularly for the initial state of the proton and its interaction with other nuclei. Another potential reason is missing physics in modeling collectivity. In the next section, it is shown that \texttt{CoMBolt-hyb} generates nearly the same collectivity as in OO collisions for systems with the same multiplicities. One should note that p--Pb events with the highest multiplicities are almost in the region where the model and data no longer agree.

\section{Opacity and scaling}\label{sec:OpacityScaling}

In Refs.~\cite{Kurkela:2018qeb,Kurkela:2019kip}, a new quantity called opacity is introduced to quantify how much the dynamic of a system is close to the ideal hydrodynamics or it behaves like a medium with few interaction among the underlying degrees of freedom. Conceptually, it is defined as the system size in the unit of the medium microscopic length-scale. Assuming the initial system size is $R$ and the microscopic length-scale is $\tau^0_{\text{relax}}$, the opacity parameter is defined as
\begin{equation}\label{eq:opacity}
	\hat{\gamma} = (R/\tau^0_{\text{relax}})^{3/4}.
\end{equation} One notes that opacity parameter is an initial state quantity in which the information of initial geometry and initial medium property are combined.

The relation to the initial state becomes more clear as follows: starting from $\tau_0 = 0^+$ and energy density $\epsilon_0$, the system requires $\tau_*=\tau^0_{\text{relax}}$ time to have at least one interaction on average, meaning the system evolves as free-streaming before the first interaction and initial energy will be $\epsilon_* = \epsilon_0 \tau_0 / \tau_*$ for a longitudinally expanding medium at the time of onset of collectivity or transverse expansion.  On the other hand, the magnitude of relaxation time at the onset of collectivity is $\tau^0_{\text{relax}} = 1/(\gamma \epsilon_*^{1/4})$. At high energies and when the system is initiated at early times system is almost scale-invariant and $\gamma$ should be a constant. By substituting $\epsilon_*$ and $\tau_*$, one can finds the initial relaxation time as $\tau^0_{\text{relax}} = \gamma^{-4/3} (\epsilon_0 \tau_0)^{-1/3}$. With this estimation, one finds the $\hat{\gamma} = \gamma R^{3/4} (\epsilon_0 \tau_0)^{1/4}$ as introduced in Refs.~\cite{Kurkela:2019kip}.

\subsection{Opacity definition for finite mass quasiparticles and at the presence of temperature dependence $\hat{H}$ }

\begin{figure*}[t]
	\centering
	\begin{tabular}{cc}
		\includegraphics[width=0.5\textwidth]{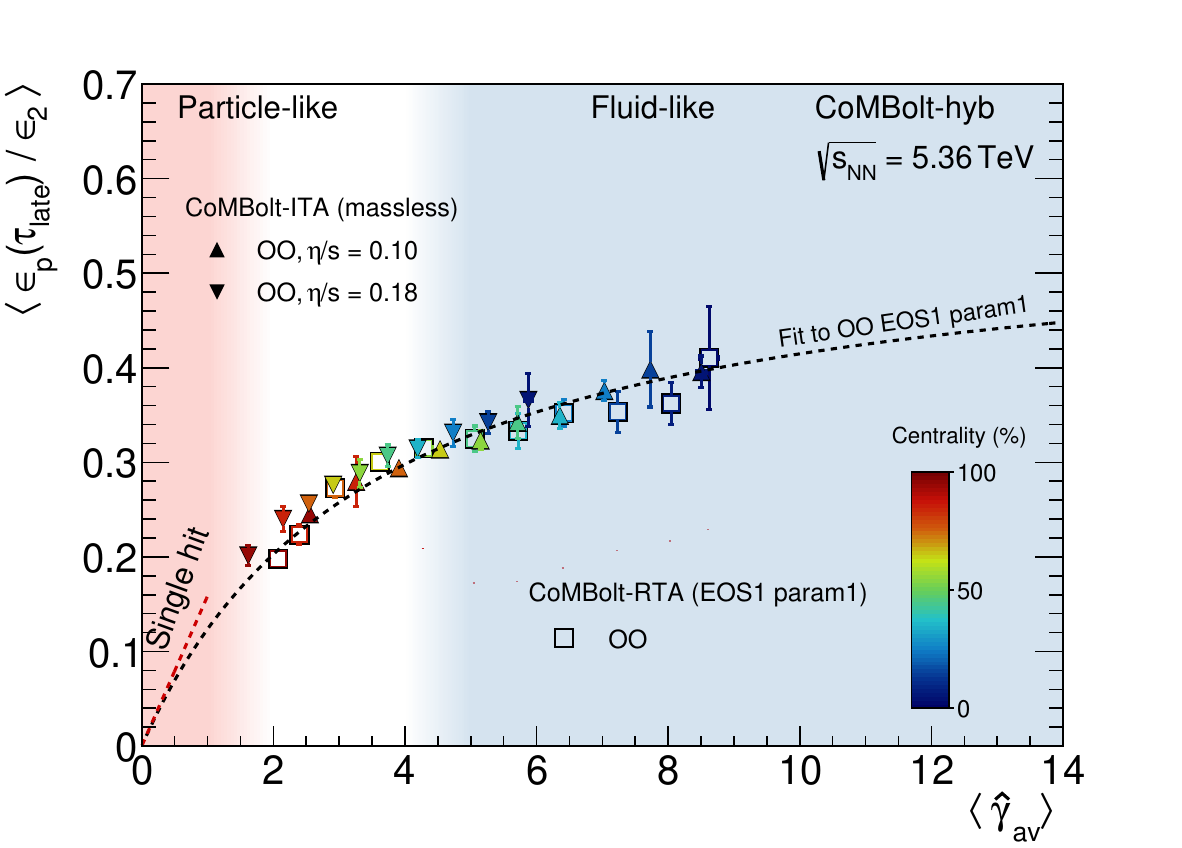} &
		\includegraphics[width=0.5\textwidth]{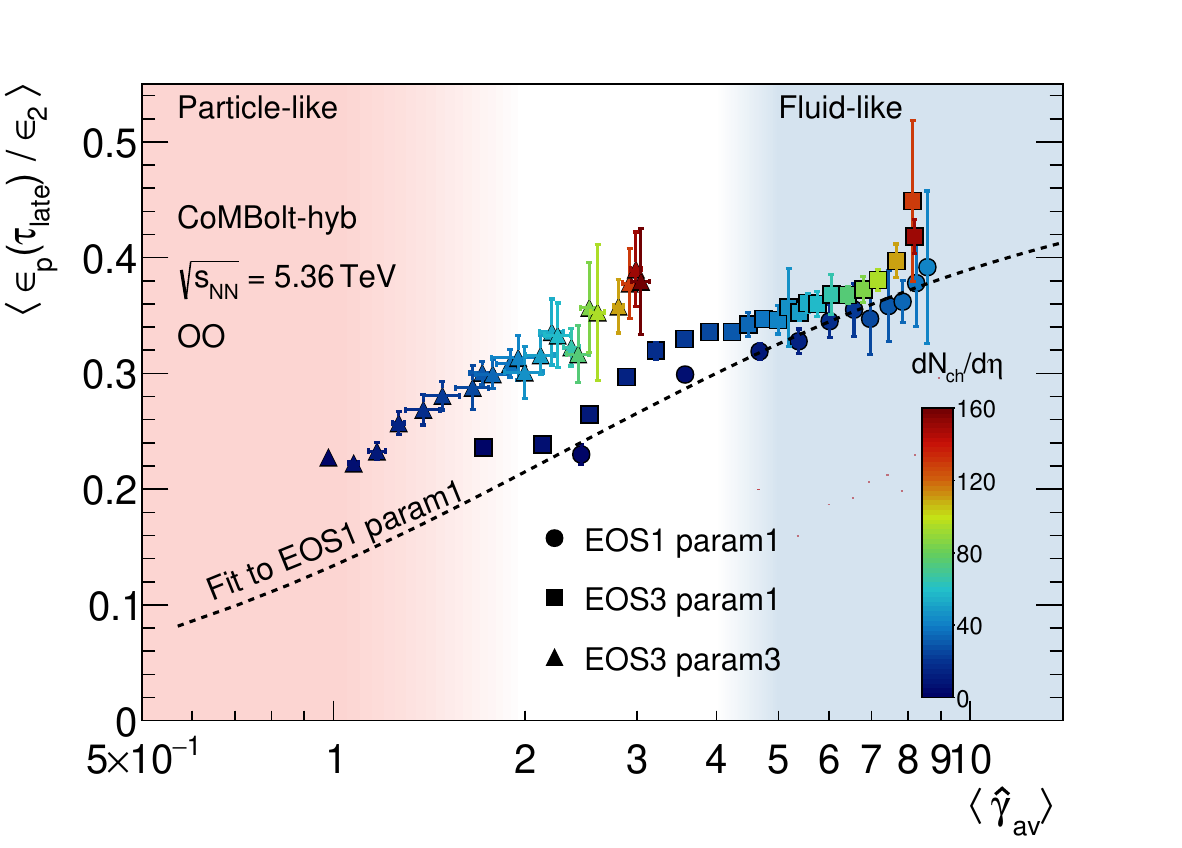} 
	\end{tabular}
\caption{Collective response as a function of opacity for OO collisions. The results obtained using the new version of \texttt{CoMBolt-hyb} with a small quasiparticle mass and constant $\hat{H}$ are compared with those obtained in the previous study in Ref.~\cite{Taghavi:2025ddm} (left). Comparison of different parameter tunings for OO collisions (right).}

	\label{fig:response-opacity}
\end{figure*}

Given that the local quantities in the transverse direction has been used in defining $\hat{\gamma}$ and that in nonconformal case $\gamma$ has spatial dependence, it is better to define average opacity from a few steps back. The final result should back to the standard definition for systems with conformal symmetry. For that, one may start from
\begin{equation}\label{eq:genericOpacity}
\hat{\gamma}_\text{av} = \left[\;\overline{\left( R/\tau^0_{\text{relax}}\right)^3}\;\right]^{1/4},
\end{equation} 
where a transverse-area normalized average is defined by
\begin{equation}
	\overline{\mathcal{O}} = \frac{1}{A_\perp} \int d^2x_\perp \mathcal{O},\quad  A_\perp = \pi R^2 
\end{equation}
and $ R^2 =  \{ x^2 + y^2\}$ is the characteristic transverse size of the system.

One could adopt a different convention for defining the opacity and may argue that conceptually the simplest definition is $\overline{R/\tau^0_{\text{relax}}}$ with no extra powers or maybe $\left[\;\overline{\left( R/\tau^0_{\text{relax}}\right)^3}\;\right]^{1/3}$. The power 1/4 in Eq.~\eqref{eq:genericOpacity} is inherited from power 3/4 in Eq.~\eqref{eq:opacity} which is the convention of Refs.~\cite{Kurkela:2018qeb,Kurkela:2019kip} since then opacity is inversely proportional to $\eta/s$ in massless limit when it is written in terms of the initial transverse energy (see Eqs.~\eqref{eq:GeneralGammaHatAv} and \eqref{eq:GeneralGammaAv}). One could have chosen other conventions, for instance in Ref.~\cite{Arslandok:2023utm} the power $3/4$ is replaced by 2/3 because then the opacity is inversely proportional to $\eta/s$ when its written in terms of transverse entropy or estimated final charged multiplicity.

The advantage of power three inside the transverse-area normalized average in Eq.~\eqref{eq:genericOpacity} is that $(\tau_{\text{relax}}^0 )^{-3}\propto \epsilon_0 \tau_0 $ and the spatial average turns into the energy weighted average.  Using $\tau^0_{\text{relax}} = \gamma^{-4/3} (\epsilon_0 \tau_0)^{-1/3}$, one obtains
\begin{equation}\label{eq:GeneralGammaHatAv}
\hat{\gamma}_\text{av} = \left[ \{ \gamma^4 \}  \frac{R}{\pi} \frac{dE_T^0}{d\eta}   \right]^{1/4},
\end{equation}
where
\begin{equation}\label{eq:GeneralGammaAv}
	\gamma(x_\perp) = \frac{1}{5 \hat{H}(T_\text{eff}) a^{1/4}_\epsilon(T_\text{eff}) }
\end{equation}
 In massless case with conformal symmetry and constant $\eta/s$, one finds $a_\epsilon(T_\text{eff}) = a_0$ and $\hat{H}(T_\text{eff}) = \eta/s$ and subsequently obtains  $\{ \gamma^4 \}^{-1/4} = 5(\eta/s) a^{1/4}_0$,
 which leads to $\hat{\gamma}_\text{av}$ as that in Refs.~\cite{Kurkela:2018qeb,Kurkela:2019kip,Ambrus:2022koq} and also used in previous OO event-by-event study  in Ref.~\cite{Taghavi:2025ddm}. Using this definition, the dependence of the collective response on opacity can be studied for different equations of state and collision systems.

\begin{figure*}[t]
	\centering
	\begin{tabular}{cc}
		\includegraphics[width=0.5\textwidth]{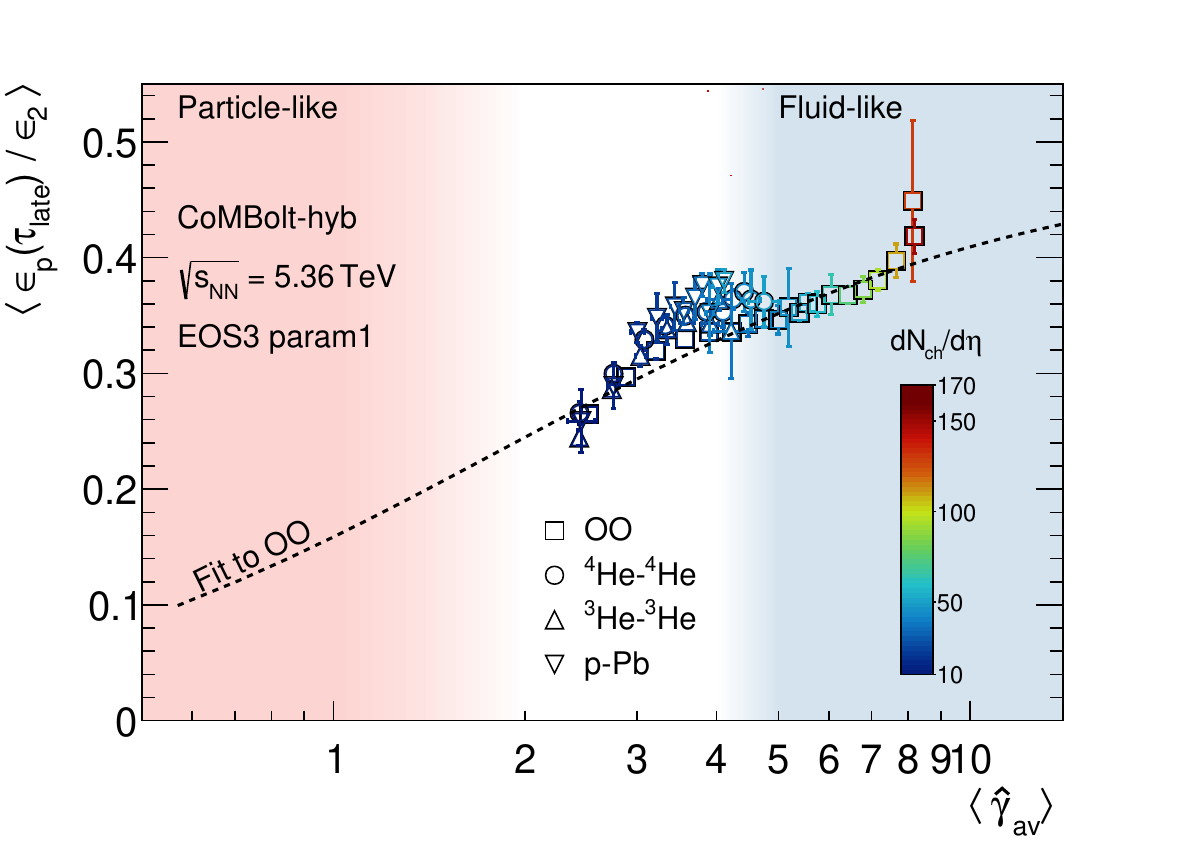} &
		\includegraphics[width=0.5\textwidth]{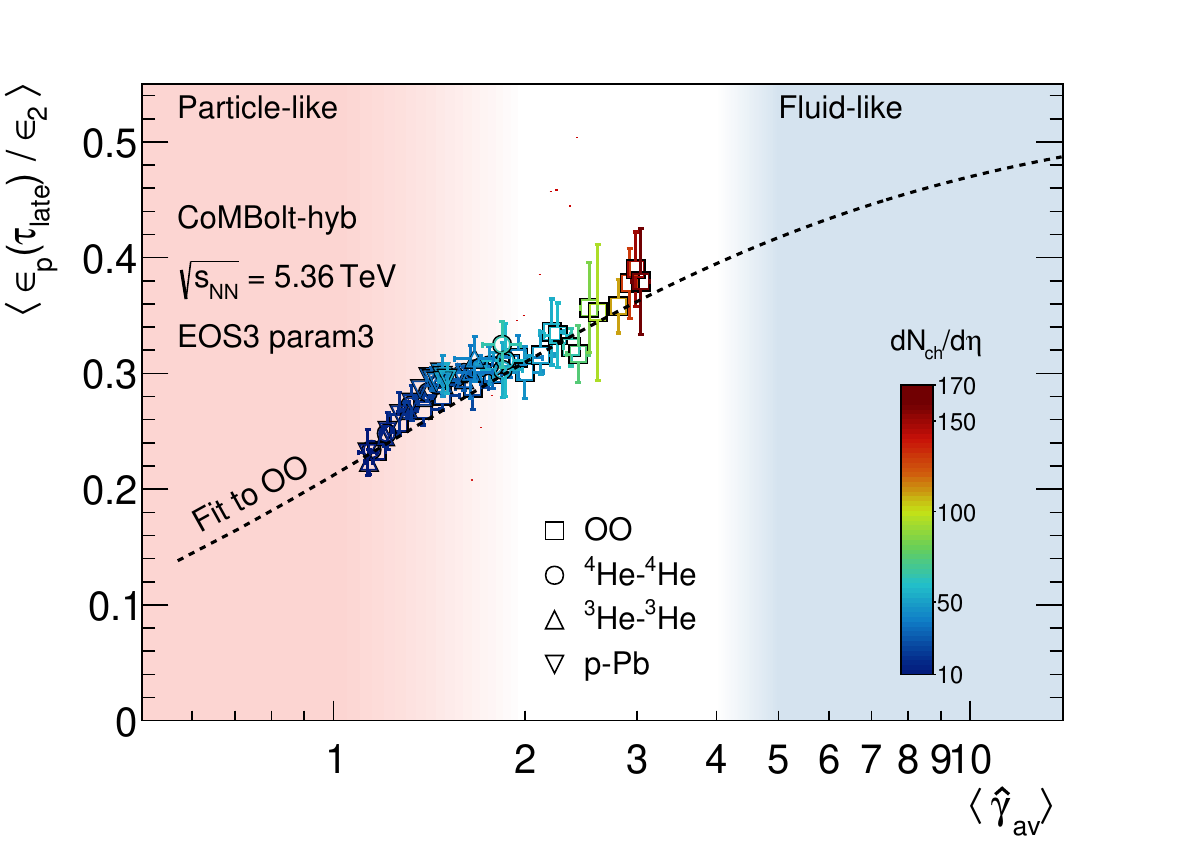} 
	\end{tabular}
\caption{Scaling properties of the collective response as a function of opacity for EOS3 with constant $\hat{H}$ (left) and temperature-dependent $\hat{H}$ (right), for four collision systems of different sizes. The parameter tuning used in the left panel is more compatible with OO collision measurements.}
	\label{fig:opacity-scaling_light_ions}
\end{figure*}

\subsection{Collectivity versus opacity}

The collective response as a function of opacity for OO collisions is shown in Fig.~\ref{fig:response-opacity} for four different parameter tunings. A measure of the level of collectivity is the momentum anisotropy defined in Eq.~\eqref{eq:momentumAnis}, divided by the initial spatial ellipticity,
\begin{equation}
	\epsilon_2 = \left| \frac{ \{ y^2-x^2 - 2i xy \} }{ \{x^2 + y^2\} } \right|.
\end{equation}
The momentum anisotropy is evaluated at $\tau_{\text{late}} = 5.0\,\text{fm}/c$, which is late enough for $\epsilon_p$ to saturate in central OO collisions (see Fig.~\ref{fig:momentAnis}). The ratio $\epsilon_p(\tau_\text{late})/\epsilon_2$ is then averaged over many events, $\la \epsilon_p(\tau_\text{late}) / \epsilon_2 \ra$. Other approaches are also possible. For instance, one could construct a two-dimensional scatter plot of $(\epsilon_2,\epsilon_p)$ and fit the relation $\epsilon_p = k_2 \epsilon_2 + k'_2 \epsilon_2^3$, where the cubic term represents the nonlinear response~\cite{Noronha-Hostler:2015dbi}. This would provide an alternative estimator of the collective response. For consistency with other studies, the first approach is adopted.

Fig.~\ref{fig:response-opacity}~(left) compares the results from the previous version of the model for massless quasiparticles in Ref.~\cite{Taghavi:2025ddm} with the EOS1 parameterization, in which the mass is rather small, $m=0.05\,$[GeV]. The param1 parameterization corresponds to $\hat{H} = 0.08$, which is close to $\eta/s = 0.1$. Other model parameters, such as the \trento{} parameters, are different. In particular, the overall normalization is much smaller for EOS1 param1 than that in Ref.~\cite{Taghavi:2025ddm}. This means that less collectivity is generated with the new tuning, and a smaller final multiplicity is produced. The latter is clear from Fig.~\ref{fig:response-opacity}~(right), where the most central collisions with EOS1 param1 produce less than half the multiplicity compared to the other tunings. However, one still observes very good agreement between the collective response as a function of opacity in the previous fully massless model and the new model with a small quasiparticle mass.

Although the collectivity as a function of opacity collapses onto a single monotonically increasing curve for different parameter tunings in the almost massless cases, this is not the case when the quasiparticle mass becomes sizable. In Fig.~\ref{fig:response-opacity}~(right), EOS1 (small mass) is compared with EOS3 (large mass). First, consider EOS1 and EOS2 with temperature-independent $\hat{H} = 0.08$. Although the curves agree at high multiplicities, a slight nonmonotonic behavior is observed toward smaller opacities.

One notes that $\epsilon/T^4$ and $P/T^4$ for massive quasiparticles depends on temperature. Moving toward smaller systems, larger regions of the system have lower temperatures, which affects the dynamics of the system. The effect becomes even more pronounced when $\hat{H}$ also becomes temperature dependent. This is seen in Fig.~\ref{fig:response-opacity}~(right) for OO collisions with EOS3 param3. It is clear from the definition that the opacity becomes smaller for larger $\hat{H}$, but at the same time, the collective response is systematically higher for this tuning. This is not the case for the massless model, as seen in Fig.~\ref{fig:response-opacity}~(left).

Despite the fact that different parameter tunings do not collapse onto a single monotonically increasing curve for massive quasiparticles, different species with different sizes still show approximate scaling. This is shown in Fig.~\ref{fig:opacity-scaling_light_ions}. It is particularly interesting to note that, for p--Pb collisions, the generated elliptic flow is much smaller than that in OO, ${}^3$He--${}^3$He, and ${}^4$He--${}^4$He collisions, as seen in Fig.~\ref{fig:v22_system_size_scan}. This suggests that the p--Pb initial state is most likely not modeled correctly and that insufficient fluctuations are introduced. However, the collective response to the initial anisotropy as a function of opacity still shows good scaling. The actual dependence of the collective response on opacity should depend on the details of the EOS and the temperature dependence of $\hat{H}$.

As discussed earlier, the best parameter tuning examined in this work is EOS3 param1. This means that Fig.~\ref{fig:opacity-scaling_light_ions}~(left) is expected to be closer to the actual opacity dependence of the collective response. The results for OO collisions are compatible with the previous study in Ref.~\cite{Taghavi:2025ddm}. In Fig.~\ref{fig:opacity-scaling_light_ions}~(right), the evolution of the system with temperature-dependent $\hat{H}$ is shown. For larger values of $\hat{H}$, the opacity of the system becomes smaller, which shifts the whole curve to the left. However, EOS3 param3 is not favored by the data. In particular, it produces too many particles, as seen in Fig.~\ref{fig:v22_charged_multi}~(right).

The advantage of OO collisions is the good theoretical understanding of their initial state. It is evident that this system is still not small enough to test particle-like dynamics in central collisions, which offer the cleanest connection to the initial nuclear structure~\cite{TabatabaeeMehr:2024lgu}. Based on the best parameter tuning found in this work, smaller systems such as ${}^3$He--${}^3$He and ${}^4$He--${}^4$He, however, lie in the transition region between fluid-like and particle-like collective dynamics, even in central collisions, where the initial state is expected to be well understood. This suggests that measuring collectivity in ${}^3$He--${}^3$He and ${}^4$He--${}^4$He collisions could provide a better probe of the transition between fluid-like and particle-like dynamics. A caveat is that the results and conclusions presented here are based on manual parameter tuning. Given the current stage of the model, a Bayesian analysis is needed to draw more robust conclusions about the transition range toward the particle-like regime~\cite{Taghavi:BayesianOO}.

\subsection{Opacity versus charged multiplicity}

The collectivity versus opacity shown in Figs.~\ref{fig:response-opacity} and \ref{fig:opacity-scaling_light_ions} is accessible only from simulations, where the initial ellipticity $\epsilon_2$ and the corresponding momentum anisotropy in the same event are available. It would be interesting to introduce a fully experimental measure of the collective response presented in these figures. There have been attempts in this direction~\cite{Ambrus:2024eqa}.

Further studies regarding a measure of the strength of the collective response are left for the future. For opacity, however, there is already a connection between opacity and final charged multiplicity~\cite{Arslandok:2023utm}. As mentioned earlier, the opacity parameter is a quantity related to the initial state. It combines the initial macroscopic size of the system with the microscopic length scale. In Eq.~\eqref{eq:GeneralGammaHatAv}, the connection between opacity and the initial energy density per unit pseudorapidity is evident. As discussed in Ref.~\cite{Giacalone:2019ldn}, one can translate the effect of the initial transverse energy at the initial time to the initial entropy at the hydrodynamization time and subsequently to the final charged multiplicity leading to $\hat{\gamma} \propto (\eta/s)^{-9/8}(dN_\text{ch}/d\eta)^{3/8}$. In a previous study, this relation was examined using the massless model, and it was observed that changing $\eta/s$ indeed changes the opacity, while the power of $dN_\text{ch}/d\eta$ is close to that estimated from idealized theoretical considerations.

\begin{figure}[t]
	\centering
	\includegraphics[width=0.5\textwidth]{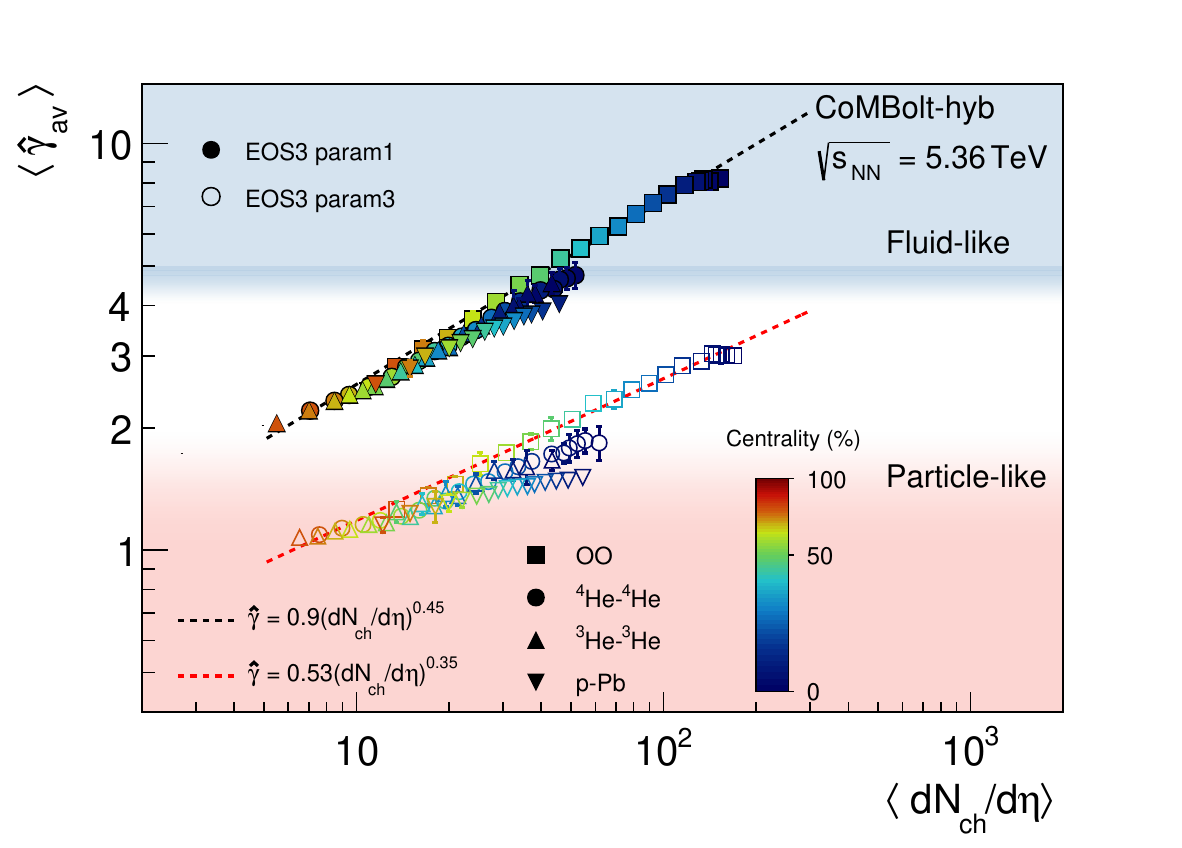}
\caption{Opacity as a function of charged multiplicity for four different collision systems and two parameter tunings. Approximate scaling is observed across the different collision systems.}
	\label{fig:opacity-multi}
\end{figure}

The opacity as a function of charged multiplicity from massive quasiparticle evolution is shown in Fig.~\ref{fig:opacity-multi}. Although approximate scaling is still observed, a slight separation between different species is seen. One notes that $\hat{H}$ approaches $\eta/s$ only in the small-mass limit, and the previous estimates should be considered only as approximations. For EOS3 param1, which is the best parameter tuning in this work, the power of $dN_\text{ch}/d\eta$ obtained by fitting the OO collision results is 0.45, which is larger than the value of 0.375 from the idealized estimate.

Although the common scaling of opacity is valid only approximately, Fig.~\ref{fig:opacity-multi} indicates that charged multiplicity, which is experimentally observable, is highly correlated with opacity. The actual correlation depends on the properties of the medium. Consequently, $(dN_\text{ch}/d\eta)^\alpha$, with $\alpha \approx 0.4$, should be a good measure of opacity, although the actual value of the exponent and the proportionality constant are model-dependent.

\section{Summary and outlook}\label{sec:conclusion}

A new version of \texttt{CoMBolt}, called \texttt{CoMBolt-RTA}, is introduced to model the evolution of massive quasiparticles in the relaxation time approximation. The model allows one to study the nonequilibrium dynamics of a nonconformal system with transverse spatial extension. The evolution of an initial Gaussian distribution from very early to late times is studied for a wide range of constant values of $\hat{H}$. In the massless limit, $\hat{H}$ approaches $\eta/s$, while at finite mass, $\hat{H}$ deviates from $\eta/s$, and the specific bulk viscosity, $\zeta/s$, becomes nonzero.

It has been shown that the anisotropy of massive quasiparticles with an initial transverse profile has rich and complicated dynamics, as demonstrated by the evolution of $(R_{\pi_L}^{-1}, R_{\pi_T}^{-1}, R_{\Pi}^{-1})$. For small values of $\hat{H}$, starting from a given point in the transverse plane, the trajectory of the inverse Reynolds numbers follows a short path toward equilibrium, while for larger values of $\hat{H}$, the trajectory follows a longer path. Due to the longitudinal expansion, at least in some regions of the medium, the system might not have enough time to reach equilibrium before it becomes very dilute and the microscopic dynamics practically stops. It has also been observed that, starting from the same initial spatial anisotropy, a system closer to free-streaming generates more momentum anisotropy for larger quasiparticle masses. However, when the system is closer to ideal hydrodynamics, the opposite behavior is observed, and a smaller quasiparticle mass generates more momentum anisotropy.

Using the hybrid version of the code, in which the simulation chain includes \trento{} for the initial state, \texttt{CoMBolt-RTA} for the partonic evolution, and \texttt{UrQMD} as an afterburner, the new model is confronted with recent OO collision measurements from ALICE and ATLAS~\cite{ALICE:2025luc,ALICE:2025ISmult,ATLAS:2026zgq}. By manually tuning the parameters, the anisotropic flow $v_2\{2\}$ and the centrality dependence of charged multiplicity are first calculated and compared with recent ALICE measurements. Including massive quasiparticles allows one to obtain an EOS closer to the lattice-QCD EOS, leading to a significant improvement in describing the $p_T$ spectrum and $\langle p_T\rangle$ compared with the massless quasiparticle model. The resulting predictions show good agreement with ATLAS measurements.

One of the motivations for developing \texttt{CoMBolt} is that it is applicable to a wide range of underlying dynamics, from fluid-like to particle-like behavior. The conclusion of the previous study in Ref.~\cite{Taghavi:2025ddm} was that central OO collisions are well within the fluid-like regime. The new version of the code, with massive quasiparticles and a new parameter tuning, does not change this conclusion. Smaller collision systems, however, might move toward a regime with more particle-like dynamics. In particular, the possibility of ${}^4$He--${}^4$He collisions in future LHC runs is under investigation. In the current study, the collectivity of ${}^3$He--${}^3$He and ${}^4$He--${}^4$He collisions is also studied. Based on the current status of the model and its parameter tuning, central ${}^4$He--${}^4$He collisions are in the transition region between fluid-like and particle-like dynamics. This is particularly interesting because the initial state of central ${}^4$He--${}^4$He collisions should be as well understood as that of central OO collisions, but the system is now around four times smaller in terms of mass number, making it a good candidate for studying the transition in collective dynamics.

There are different directions one could follow to develop the model further. A more systematic parameter tuning based on Bayesian techniques is ongoing, which should lead to more robust parameter estimation~\cite{Taghavi:BayesianOO}. As discussed earlier, assuming a constant quasiparticle mass does not provide the full freedom to choose an input EOS, as is possible in conventional hydrodynamic codes. Introducing a temperature-dependent mass would therefore be an important future improvement. Furthermore, the relaxation time approximation provides a simplified collision kernel. A more realistic collision kernel based on effective kinetic theory (EKT) could be employed in the future, particularly in high-temperature regions~\cite{Arnold:2002zm}.

\section{Acknowledgment}

The author had fruitful discussions with the participants of the ``EMMI Workshop: Light-Ion Collisions 2026'' in Heidelberg, where some of the results of this work were presented for the first time.
AI tools were used to assist with plot preparation, grammar check, and minor debugging of the CoMBolt code. All code modifications and resulting outputs were checked and validated by the author, who takes full responsibility for the results presented in this work. The author is supported by the Deutsche Forschungsgemeinschaft (DFG) through grant number 517518417.

\bibliographystyle{unsrt}  
\bibliography{references}%

\end{document}